\documentclass{amsproc}

\usepackage{amssymb}

\newcommand{\im}{\mathrm{Im\,}}
\newcommand{\e}{\mathrm{e}}
\newcommand{\D}{\mathrm{d}}
\newcommand{\C}{\mathbb{C}}
\newcommand{\N}{\mathbb{N}}
\newcommand{\R}{\mathbb{R}}
\newcommand{\Z}{\mathbb{Z}}
\newcommand{\BB}{\mathcal{B}}
\newcommand{\CC}{\mathcal{C}}
\newcommand{\DD}{\mathcal{D}}
\newcommand{\EE}{\mathcal{E}}
\newcommand{\OO}{\mathcal{O}}
\newcommand{\PP}{\mathcal{P}}
\newcommand{\QQ}{\mathcal{Q}}
\newcommand{\RR}{\mathcal{R}}
\newcommand{\SSS}{\mathcal{S}}
\newcommand{\TT}{\mathcal{T}}

\newtheorem{theorem}{Theorem}[section]

\newtheorem{proposition}[theorem]{Proposition}
\newtheorem{corollary}[theorem]{Corollary}

\theoremstyle{definition}

\theoremstyle{remark}
\newtheorem{remark}[theorem]{Remark}
\newtheorem{remarks}[theorem]{Remarks}
\numberwithin{equation}{section}

\begin{document}

\title[Leaky Quantum Graphs]{Leaky Quantum Graphs: A Review}

\author{Pavel Exner}
\address{Department of Theoretical Physics, Nuclear Physics
Institute, Czech Academy of Sciences, 25068 \v{R}e\v{z} near Prague, and
Doppler Institute for Mathematical Physics and Applied
Mathematics, Czech Technical University, B\v{r}ehov\'{a}~7, 11519
Prague, Czechia} \email{exner@ujf.cas.cz}
\thanks{The research was supported in part by the Czech Academy
of Sciences and Ministry of Education, Youth and Sports within the
projects A100480501 and LC06002. The author thanks Peter Kuchment and the referee for comments, which helped to improve the manuscript.}

\date{September ??, 2002 and, in revised form, ????, 2002.}


\begin{abstract}
The aim of this review is to provide an overview of a recent work
concerning ``leaky'' quantum graphs described by Hamiltonians
given formally by the expression $-\Delta -\alpha \delta
(x-\Gamma)$ with a singular attractive interaction supported by a
graph-like set in $\mathbb{R}^\nu,\: \nu=2,3$. We will explain how
such singular Schr\"odinger operators can be properly defined for
different codimensions of $\Gamma$. Furthermore, we are going to
discuss their properties, in particular, the way in which the
geometry of $\Gamma$ influences their spectra and the scattering,
strong-coupling asymptotic behavior, and a discrete counterpart to
leaky-graph Hamiltonians using point interactions. The subject
cannot be regarded as closed at present, and we will add a list of
open problems hoping that the reader will take some of them as a
challenge.
\end{abstract}

\maketitle

\tableofcontents

\section{Introduction}

In this paper we are going to review results concerning a class of
``different'' quantum graph models. With this aim in mind, it
would be natural to start by recalling briefly the standard
quantum graphs, their description, properties, and numerous
applications. In this volume, however, this would be clearly to
bring owls to Athens\footnote{Or, depending on your taste, coal to
Newcastle, firewood to the forest, etc. As usual, one can also
refer to the Bard: \emph{to throw a perfume on the violet} (The
Life and Death of King John).} and we refrain from doing that
referring to the other articles in these proceedings, or to
\cite{BCFK} as another rich bibliography source.

To motivate a need to look for alternative description of
graph-like structures, let us observe that --- despite its
mathematical simplicity, beauty, and versatility --- the standard
quantum-graph model has also some drawbacks. In our opinion, the
following two are the most important:
\begin{itemize}
\item the presence of \emph{ad hoc} parameters in the boundary
conditions which describe how the wave functions are matched at
the graphs vertices
\item the fact that particles are strictly confined to graph
edges. While this is often a reasonable approximation when
dealing, say, with electrons in semiconductor quantum wires, such
a model neglects \emph{quantum tunneling} which can play role once
such wires are placed close to each other. One consequence is that
in such a description, with the graph embedded in
$\mathbb{R}^\nu$, spectral properties reflect the topology while
the graph geometry enters only through the edge lengths, their
shapes being irrelevant
\end{itemize}
A way to deal with the first problem is to regard a quantum graph
as an idealization of a more ``realistic'' system without such
ambiguities; a natural candidate for this role are various ``fat
graphs''. Limits of such objects when the edge width squeezes to
zero were studied extensively, first in the easier Neumann-type
case \cite{FW, KZ, RS, Sa, Ku2, KZ2, EP1, Po1, EP2} and quite
recently also in the situation with Dirichlet boundaries
\cite{Po2, MV, CaE, Gr}. These results give a partial answer to
the first question\footnote{An alternative approach is to keep the
graph fixed and to approximate the vertex coupling through
suitably scaled families of regular or singular interactions --
see \cite{Ex1, ChE, ET}.} while the second problem remains.

Here we are going to discuss a class of quantum graph models which
are free of both difficulties; the price we pay is --- similarly
as for the fat graphs mentioned above --- that instead of
ordinary differential equations we have to deal with a PDE
problem. The idea is to preserve the whole Euclidean space as the
configuration space and to suppose that the particle is kept in
the vicinity of the graph $\Gamma\subset \mathbb{R}^\nu$ by an
attractive singular interaction. Formally such a Hamiltonian
expresses as
\begin{equation} \label{formal}
-\Delta -\alpha(x) \delta (x-\Gamma )\,
\end{equation}
with $\alpha(x)>0\,$; we will consider mostly the situation where
the attraction is position independent, $\alpha(x)=\alpha>0$.
Before proceeding to a definition of such singular Schr\"odinger
operators and discussion of their properties, let us make a few
remarks.

First of all, it is clear that there is no ambiguity related to
the graph vertices once $\Gamma$ and $\alpha$ are given. It is
equally obvious that  the confinement in this model takes place at
negative energies only. The particle now ``lives'' in the whole
space and can be found even at large distances from $\Gamma$,
although with a small probability, because the complement
$\mathbb{R}^\nu \setminus\Gamma$ is the classically forbidden
region. The presence of the tunneling is the reason why we dub
such systems as \emph{leaky quantum graphs}.

Schr\"odinger operators with interactions supported by curves and
other manifolds of a lower dimension were studied already in the
early nineties \cite{BT}, and even earlier in examples with a
particular symmetry \cite{AGS, Sha}. A more systematic
investigation motivated by the above considerations was undertaken
in a last few years; it is the aim of this review to describe its
results.

One should stress, however, that such mathematical structures can
be studied also from other points of view. A prominent example
comes from studies of high contrast optical systems\footnote{Another situation where one arrives at a leaky-graph-type model arises when one deals with contact interactions of several one-dimensional particles \cite{Du}.} used to model photonic crystals --- see, e.g. \cite{FK, KK} --- which in a suitable approximation yield an analogue of the spectral problem for the operator (\ref{formal}); the two differ only by the physical interpretation, the roles of the coupling and spectral parameters being switched. A derivation of leaky-graph models in this context was given in the paper \cite{FK2}, see also the review \cite{Ku} and recall that the corresponding operators can be cast also in a pseudo-differential form \cite{FK2, PP}. 

The material we are going to review is relatively extensive. We
will take care, of course, to explain properly all the notions and
the results. On the other hand, proofs will be mostly sketched.
However, we will always give references to original papers where
the particular complete argument can be found. Let us finally
remark that the subject reviewed here cannot be regarded as
closed, on the contrary, there are many open questions. We devote
to them the closing section, and the author of this survey can
only hope that his reader will take this problem list as a
challenge and a program which will keep him or her busy for some
time.

\section{Leaky graph Hamiltonians}

\subsection{Quadratic forms and boundary conditions} \label{ss:
qf&bc}

The Hamiltonians we are interested in are generalized
Schr\"odinger operators with a singular interaction supported by a
graph-like $\Gamma$ which is a zero measure set in $\R^\nu$. We
will use facts about such operators derived, in particular, in
\cite{BEKS} specifying them to our present purpose. Let us first
suppose that the configuration space dimension $\nu=2$ and the
coupling ``strength'' is constant on the interaction support.

To begin with, let us show how such a singular operator can be
defined generally through the associated quadratic form. Consider
a positive Radon measure $m$ on $\R^2$ and a number $\alpha>0$
such that
\begin{equation} \label{basiccond}
(1+\alpha) \int_{\R^2} |\psi(x)|^2\, \D m(x) \le a \int_{\R^2}
|\nabla\psi(x)|^2\, \D x + b \int_{\R^2} |\psi(x)|^2\, \D x
\end{equation}
holds for all $\psi\in\SSS(\R^2)$ and some $a<1$ and $b$. The map
$I_m$ defined by $I_m\psi=\psi$ on the Schwartz space $\SSS(\R^2)$
extends by density uniquely to
\begin{equation}
I_m:\: W^{1,2}(\R^2) \,\to\, L^2(m):= L^2(\R^2,m) \;;
\end{equation}
for brevity the same symbol is used for a continuous function and
the corresponding equivalence classes in both $L^2(\R^2)$ and
$L^2(m)$. The inequality (\ref{basiccond}) extends to
$W^{1,2}(\R^2)$ with $\psi$ replaced by $I_m\psi$ at the left-hand
side. The quadratic form
\begin{equation} \label{Hamform}
\EE_{-\alpha m}[\psi] :=  \int_{\R^2} |\nabla\psi(x)|^2 \, \D x
-\alpha \int_{\R^2} |(I_m\psi)(x)|^2\, \D m(x)
\end{equation}
is defined on $W^{1,2}(\R^2)$; it is straightforward to check
\cite{BEKS} that under the condition (\ref{basiccond}) this form
is closed and below bounded, with $C_0^{\infty}(\R^2)$ as a core,
and consequently, it is associated with a unique self-adjoint
operator denoted as $\hat H_{-\alpha m}$. A sufficient condition
for the inequality (\ref{basiccond}) to be valid is that the
measure $m$ belongs to the generalized Kato class, i.e.
\begin{equation} \label{Kato}
\lim_{\epsilon\to 0}\: \sup_{x\in\R^2}\, \int_{B_\epsilon(x)} |\ln
|x\!-\!y|| \, \D m(y) = 0\,,
\end{equation}
where $B_\epsilon(x)$ is the ball of radius $\epsilon$ centered at
$x$. In such a case, moreover, any positive number can be  chosen
as $a$. So far the construction has been general and involved also
regular Schr\"odinger operators. Suppose now that $m$ is the Dirac
measure supported by a graph $\Gamma\subset\R^2$ which has the
following properties:
\begin{description}
\item[\emph{(g1)} edge smoothness] each edge $e_j\in\Gamma$ is a
graph of $C^1$ function $\gamma_j:\:I_j\to\R^2$ where $I_j$ is an
interval (finite, semi-infinite, or the whole $\R$). Moreover,
without loss of generality we may suppose that edges are
para\-metrized by the arc length, $|\dot\gamma_j(s)|=1$.
\item[\emph{(g2)} cusp absence] at the vertices of $\Gamma$ the
edges meet at nonzero angles.
\item[\emph{(g3)} local finiteness] each compact subset of $\R^2$
contains at most a finite number of edges and vertices of
$\Gamma$.
\end{description}
The last assumption allows us to extend Theorem~4.1 of \cite{BEKS}
applying it to the Dirac measure supported by the graph. More
exactly, we consider the measure
\begin{equation} \label{meas}
m_\Gamma:\; m_\Gamma(M) = \ell_1(M\cap\Gamma)
\end{equation}
for any Borel $M\subset\R^2$, where $\ell_1$ is the
one-dimensional Hausdorff measure given in our case by the
edge-arc length. Such a straightforward extension implies that
$m\equiv m_\Gamma$ satisfies the condition (\ref{basiccond}) and
gives thus rise to the appropriate operator $\hat H_{-\alpha m}$;
to make the notation explicit we will employ for it in the
following the symbol $H_{\alpha,\Gamma}$. This is one way how to
give meaning to the formal expression (\ref{formal}).

An alternative is to use boundary conditions. Consider the
operator acting as
$$ \left(\dot H_{\alpha, \Gamma}\psi \right)(x) =
-(\Delta\psi)(x)\,, \quad x\in \R^2\setminus\Gamma\,, $$
on any function $\psi$ which belongs to
$W^{2,2}(\R^2\setminus\Gamma)$, is continuous at each edge
$e_j\in\Gamma$ with the normal derivatives having there a jump,
namely
\begin{equation} \label{2-1bc}
 {\partial\psi\over\partial n_+}(x) - {\partial\psi\over\partial
n_-}(x) = -\alpha \psi(x)\,, \quad x\in \mathrm{int\,}e_j\,;
\end{equation}
since the edges are smooth by assumption, the normal vector exists
at each inner point of an edge. In the same way as in \cite{BEKS}
one can check that $\dot H_{\alpha,\Gamma}$ is e.s.a., and
moreover, by Green's formula it reproduces the form
(\ref{Hamform}) on its core, so its closure may be identified with
$H_{\alpha,\Gamma}$ defined above.

  \begin{remarks} \label{rem1def}
  (i) The above definitions easily extend to the situation with the
  singular interaction strength $\alpha(s)$ varying along the edges provided
  the corresponding function $\alpha:\Gamma\to\R_+$ is sufficiently
  regular. \\
  (ii) In a similar way one can define operators corresponding to
  the formal expression (\ref{formal}) for a generalized ``graph''
  whose edges are $(\nu-1)$--dimensional manifolds in $\R^\nu$
  satisfying suitable regularity conditions analogous to
  (g1)--(g3).
  \end{remarks}

\subsection{Regular potential approximation} \label{ss: regular}

As we shall see below the operators $H_{\alpha,\Gamma}$ represent
a reasonably general class of systems for which various properties
can be derived. One can ask nevertheless whether this is not again
a too idealized model. Before proceeding further we want to show
that $H_{\alpha,\Gamma}$ can be regarded as weak-coupling
approximation to a class of regular Schr\"odinger operators; for
simplicity we restrict ourself to graphs with a single infinite
edge.

Let $\Gamma$ be a curve described by a $C^2$ function
$\gamma:\:\R\to\R^2$. Then we are able to define the signed
curvature $k(s):= \left(\dot\gamma_1 \ddot\gamma_2 - \dot\gamma_2
\ddot\gamma_1 \right)(s)$; we assume that it is bounded,
$|k(s)|<c_+$ for some $c_+>0$ and all $s\in\R$. Moreover, we
suppose that $\Gamma$ has neither self-intersections nor
``near-intersections'', i.e. that there is a $c_->0$ such that
$|\gamma(s)\!-\!\gamma(s')|\ge c_-$ for any $s,s'$ with
$|s\!-\!s'|\ge c_-$. Then we can define in the vicinity of
$\Gamma$ the standard locally orthogonal system of coordinates
\cite{ES}, i.e. the pairs $(s,u)$ where $u$ is the (signed) normal
distance from $\Gamma$ and $s$ is the arc-length coordinate of the
point of $\Gamma$ where the normal $n(s)$ is taken; the system is
unique in the strip neighborhood $\Sigma_{\epsilon}:= \{ x(s,u):\:
(s,u)\in \Sigma^0_{\epsilon}\}$, where
\begin{equation} \label{curvi}
x(s,u):= \gamma(s) + n(s)u
\end{equation}
and $\Sigma^0_{\epsilon}:= \{(s,u):\: s\in\R, |u|<\epsilon\}$ as
long as the condition $2\epsilon< c_-$ is valid.

With these prerequisites we can construct the approximating
family. Given $W\in L^{\infty}((-1,1))$, we define for all
$\epsilon< {1\over 2}\, c_-$ the transversally scaled potential,
\begin{equation} \label{scaledpot}
V_{\epsilon}(x) := \left\lbrace
\begin{array}{ccl} 0 & \quad\dots\quad & x\not\in\Sigma_{\epsilon}
\\ [.5em] -\,{1\over\epsilon}\, W\left(u\over\epsilon \right) &
\quad\dots\quad & x\in\Sigma_{\epsilon} \end{array} \right.
\end{equation}
and put
\begin{equation} \label{scaledHam}
H_{\epsilon}(W,\gamma):= -\Delta\, +\, V_{\epsilon} \,.
\end{equation}
The operators $H_{\epsilon}(W,\gamma)$ are obviously self-adjoint
on $D(-\Delta)= W^{2,2}(\R^2)$ and we have the following
approximation result:
\begin{theorem} \label{ditch_approx}
Under the stated assumptions, $H_{\epsilon}(W,\Gamma) \to
H_{\alpha, \Gamma}$ as $\epsilon\to 0$ in the norm-resolvent
sense, where $\alpha := \int_{-1}^1 W(t)\, \D t$.
\end{theorem}
\begin{proof}[Sketch of the proof] One has to compare the resolvents,
that of $H_{\alpha, \Gamma}$ given below and the Birman-Schwinger
expression of $\left(H_{\epsilon}(W,\gamma) \!-\!k^2
\right)^{-1}$. Both are explicit integral operators and their
difference can be treated in a way similar to that used in the
squeezing approximation of the one-dimensional $\delta$
interaction --- see, e.g., \cite{AGHH} --- a full account of the
argument can be found in \cite{EI}.
\end{proof}

Notice that the regular potential approximation is not the only
way how to justify the leaky-graph model physically; for an
alternative see \cite{FK2}.

\subsection{The resolvent}

As usual, the spectral and scattering properties are encoded in
the resolvent and our first task is to find an explicit expression
for this operator. We will employ an analogue of the
Birman-Schwinger formula for our singular case. If $k^2$ belongs
to the resolvent set of $H_{\alpha,\Gamma}$ we put
$R^k_{\alpha,\Gamma} := (H_{\alpha,\Gamma}-k^2)^{-1}$. We look for
the difference of this operator and the free resolvent $R^k_0$
which is for $\im k>0$ an integral operator with the kernel
\begin{equation} \label{freeG}
G_k(x\!-\!y) = {i\over 4}\, H_0^{(1)} (k|x\!-\!y|)\,.
\end{equation}
To this aim we need embedding operators associated with $R^k_0$.
Let $\mu, \nu$ be arbitrary positive Radon measures on $\R^2$ with
$\mu(x)= \nu(x) =0$ for any $x\in\R^2$. By $R^k_{\nu,\mu}$ we
denote the integral operator from $L^2(\mu):=L^2(\R^2,\mu)$ to
$L^2(\nu)$ with the kernel $G_k$, i.e.
$$ R^k_{\nu,\mu} \phi = G_k \ast \phi\mu $$
holds $\nu$-a.e. for all $\phi\in D(R^k_{\nu,\mu}) \subset
L^2(\mu)$. In our case the two measures will be $m\equiv m_\Gamma$
introduced by (\ref{meas}) and the Lebesgue measure $\D x$ on
$\R^2$ in different combinations, which simply means that one or
both variables in the kernel (\ref{freeG}) are restricted to
$\Gamma$. Using this notation we can state the following result:
\begin{proposition} \label{BS}
(i) There is a $\kappa_0>0$ such that the operator $I-\alpha
R^{i\kappa}_{m,m}$ on $L^2(m)$ has a bounded inverse for any
$\kappa \ge \kappa_0$. \\ [1mm]
(ii) Let $\im k>0$. Suppose that $I-\alpha R^k_{m,m}$ is
invertible and the operator
$$ R^k := R_0^k + \alpha R^k_{\D x,m} [I-\alpha R^k_{m,m}]^{-1}
R^k_{m,\D x} $$
from $L^2(\R^2)$ to $L^2(\R^2)$ is everywhere defined. Then $k^2$
belongs to $\rho(H_{\alpha,\Gamma})$ and
$(H_{\alpha,\Gamma}-k^2)^{-1}= R^k$.
\\ [1mm]
(iii) $\:\dim\ker(H_{\alpha,\Gamma}-k^2) = \dim\ker(I-\alpha
R^k_{m,m})$ for any $k$ with $\im k>0$. \\ [1mm]
(iv) an eigenfunction of $H_{\alpha,\Gamma}$ associated with such
an eigenvalue $k^2$ expresses as
 $$ \psi(x)= \int_0^L R^k_{\D x,m}(x,s) \phi(s)\, \D s\,, $$
where $\phi$ is the corresponding eigenfunction of $\alpha
R^k_{m,m}$ with the eigenvalue one.
\end{proposition}
\begin{proof}[Sketch of the proof]
The result, which is in fact valid for any operator $\hat
H_{-\alpha m}$ of Sec.~\ref{ss: qf&bc}, is obtained by verifying
the Birman-Schwinger (BS) formula in our singular setting. The
procedure requires some care; a full account concerning the claims
(i)--(iii) can be found in \cite{BEKS}, for (iv) see \cite{Pos2}.
\end{proof}

\subsection{The case of codimension two} \label{ss: cod2}

The second one of the above constructions of $H_{\alpha,\Gamma}$
can also be rephrased in the following way: we first restrict the
Laplacian to a symmetric operator defined on function which vanish
in the vicinity of $\Gamma$, and afterwards we choose a particular
self-adjoint extension specified by the condition (\ref{2-1bc}).
It follows from general properties of partial differential
operators \cite{He} that a similar construction is possible also
in higher dimensions as long as $\mathrm{codim\,}\Gamma\le 3$.
Since we want to stick to cases of physical interest, we will
mention here only graphs whose edges are curves in $\R^3$.

An analogue of the form definition (\ref{Hamform}) does not work
in this situation and we have to rely on boundary conditions which
are, however, more complicated than (\ref{2-1bc}). The difference
is of a local character, thus we restrict ourself to the simplest
situation when $\Gamma$ is a single infinite curve described by a
$C^2$ function $\gamma (s):\:\R\to \R^{3}$ without
self-intersections and such that $|\dot \gamma(s)|=1$. In view of
the smoothness assumption the curve possesses locally Frenet's
frame, i.e. the triple $(t(s),b(s),n(s))$ of the tangent, binormal
and normal vectors; we assume its global existence\footnote{This
is true, for instance, if $\ddot\gamma$ does not vanish. For
curves having isolated straight segments a suitable coordinate
system can be obtained by patching local Frenet systems together,
possibly with a rotation -- cf.~a discussion on that point in
\cite{EK3}.}. For a fixed nonzero $\rho \in \R^2$ we define the
``shifted'' curve $\Gamma_{\rho}$ as the graph of the function
 $$
 \gamma_{\rho}(s)):=\gamma(s)+\rho_1 b(s)+\rho_2 n(s)\,;
 $$
the distance between the two is obviously $r:=|\rho|$. If we
suppose in addition that $\Gamma$ does not have
``near-intersections'' as in Sec.~\ref{ss: regular} then clearly
$\Gamma_{\rho}\cap \Gamma =\emptyset$ holds provided $r$ is small
enough. Since any function $f\in W_{loc}^{2,2}(\R^{3}\setminus
\Gamma )$ is continuous on $\R^{3}\setminus \Gamma$ its
restriction to $\Gamma_{\rho}$ is then well defined; we denote it
as ${f\big|}_{\Gamma_{\rho}}$, in fact, we can regard
${f\big|}_{\Gamma_{\rho}}$ as a distribution from $D^{\prime}(\R)$
with the parameter $\rho$. We denote by $\DD$ the set of functions
$f\in W_{loc}^{2,2}(\R^{3}\setminus \Gamma )\cap L^{2}(\R^{3})$
such that the following limits
 \begin{eqnarray*}
 \Xi (f)(s) &\!:=\!& -\lim_{r \to 0}\frac{1}{\ln r }
 {f\big|}_{\Gamma_{\rho}}(s)\,, \\
 \Omega (f)(s) &\!:=\!& \lim_{r \to 0}
 {f\big|}_{\Gamma_{\rho}} (s)+\Xi (f)(s)\ln r\,,
 \end{eqnarray*}
exist a.e. in $\R$, are independent of the direction ${1\over
r}\rho$, and define functions from $ L^{2}(\R)$; the limits here
are understood in the sense of the $D^{\prime }(\R)$ topology. Now
we are able to define the singular Schr\"odinger operator in the
present case: it acts as
\begin{equation} \label{cod2Ham}
H_{\alpha,\Gamma}f = -\Delta f \quad\mathrm{for}
 \quad x \in \R^{3}\setminus \Gamma
\end{equation}
on the domain
\begin{equation} \label{cod2bc}
D(H_{\alpha,\Gamma}) := \{\,g\in \DD :\:2\pi \alpha \Xi
 (g)(s)=\Omega (g)(s)\}\,.
\end{equation}
In this way we get a well-defined Hamiltonian which we seek
\cite{EK2}:
\begin{theorem}  \label{cod2sa+res}
Under the stated assumptions the operator $H_{\alpha,\Gamma}$ is
self-adjoint.
\end{theorem}
As in the case of codimension one and the conditions (\ref{2-1bc})
the above definition has a natural meaning as a point interaction
in the normal plane to $\Gamma$.

The proof of Theorem~\ref{cod2sa+res} is technically more involved
and we restrict ourselves to a few remarks referring to \cite{EK2}
for the full exposition. The argument is in a sense opposite to
the previous considerations. It is based on an abstract analogue
of Proposition~\ref{BS} proved in \cite{Pos1}, see also
\cite{Pos2}, which shows existence of a self-adjoint operator with
the resolvent of the appropriate form, after that one verifies
that this operator coincides with the above $H_{\alpha,\Gamma}$.

  \begin{remark} \label{r:3BS}
Since the mentioned resolvent formula analogous to that of
Proposition~\ref{BS} will be used in the following, we will
describe it at least briefly. It contains again traces of the free
resolvent, which is now given by
\begin{equation} \label{freeG3}
G_k(x\!-\!y) = \frac{\e^{ik|x-y|}} {4\pi \left| x\!-\!y\right|}\,.
\end{equation}
However, the use of Posilicano's abstract result requires to
interpret the embedding operators involved not as maps between
$L^2$ spaces, but rather the last factor $R_{\Gamma}^k$ as
$W^{2,2}(\R^3)\to L^2(\R)$ and its counterpart
$\breve{R}_{\Gamma}^k$ as the Banach space dual to
$\overline{R_{\Gamma}^k}$. Then
 \begin{equation} \label{3BS}
 R^k= R_0^k
 -\breve{R}_{\Gamma}^k (Q^k\!-\!\alpha )^{-1} R_{\Gamma}^k\,,
 \end{equation}
where the modified position of $\alpha$ in this formula
corresponds to the usual convention about the coupling parameter
for two-dimensional point interactions \cite{AGHH} reflected in
(\ref{cod2bc}); roughly speaking, it is an inverse of the one
appearing in (\ref{2-1bc}). The operator $Q^k$ is the counterpart
to $R^k_{m,m}$ of Proposition~\ref{BS} but we use on purpose a
different symbol to stress that we cannot write it simply as an
integral operator and a renormalization is needed, cf.~\cite{EK2}
for more details.
  \end{remark}

\section{Geometrically induced properties} \label{s: geom}

We said in the opening that even if we think about $\Gamma$ in the
usual quantum graph model as embedded in $\R^\nu$, the shapes of
the edges do not influence the spectrum. Leaky graphs are
different as one can illustrate in various ways.

\subsection{Bound states due to non-straightness} \label{ss:
nonstr}

Consider again a leaky graph in $\R^2$. If $\Gamma=\Gamma_0$ is a
straight line corresponding to $\gamma_0(s)= as+b$ for some
$a,b\in\R^2$ with $|a|=1$, we can separate variables and show that
\begin{equation} \label{straightspec}
\sigma(H_{\alpha,\Gamma_0}) = \Big\lbrack - \frac14\, \alpha^2,
\infty \Big)
\end{equation}
is purely absolutely continuous. We are going to show that a bend
or deformation produces, within a wide class of curves $\Gamma$, a
non-void discrete spectrum. To be specific, we assume that the
generating function $\gamma:\:\R\to\R^2$ is continuous and
piecewise $C^1$ (or, in terms of the assumption (g1)--(g3), a
graph which may have vertices but no branchings) satisfying the
following conditions\footnote{If $\gamma\in C^2$ we have a
sufficient condition for (a2) in terms of the signed curvature
introduced in Sec.~\ref{ss: regular}: it is valid with $\mu>
\frac12$ if $k(s) =\OO(|s|^{-\beta})$ with $\beta> \frac54$ as
$|s|\to\infty$, cf.~\cite{EI}.}:
\begin{description}
 \item[{\em (a1) a lower distance bound}] there is $c\in(0,1)$ such
 that $|\gamma(s)-\gamma(s')| \ge c|s-s'|$. In particular, $\Gamma$
 has no cusps and self-intersections, and its possible asymptotes
 are not parallel to each other.
 \item[{\em (a2) asymptotic straightness}] there are positive
 $d,\, \mu>\frac12$, and $\omega\in(0,1)$ such that the inequality
 $$ 
1-\, {|\gamma(s)-\gamma(s')|\over|s-s'|} \le d \left\lbrack
1+|s+s'|^{2\mu} \right\rbrack^{-1/2}
 $$ 
 holds true in the sector $S_\omega:= \left\{ (s,s'):\: \omega <
 {s\over s'} < \omega^{-1}\, \right\}$.
 \item[{\em (a3) non-triviality}] we excluded the case
 $\Gamma=\Gamma_0$. Recall that
 $$ 
|\gamma(s)-\gamma(s')| \le |s-s'|
 $$ 
holds for any $s,s'\in \R$, hence we request in other words that
the last inequality is sharp at least for some $s,s'\in \R$.
\end{description}
Then we have the following result:
\begin{theorem} \label{dsexist}
Let $\alpha>0$ and suppose that $\gamma: \R\to\R^2$ satisfies the
above assumptions. Then the essential spectrum is the same as for
the straight line, $\sigma_\mathrm{ess}(H_{\alpha,\Gamma}) =
\left\lbrack -{1\over 4}\alpha^2, \infty \right)$, but
$H_{\alpha,\Gamma}$ has at least one isolated eigenvalue below
$-{1\over 4}\alpha^2$.
\end{theorem}
\begin{proof}[Sketch of the proof]
Observe first that in view of (a2) it is not difficult to
construct a Weyl sequence to $H_{\alpha,\Gamma}$ showing that any
non-negative number belongs to $\sigma_\mathrm{ess}$. To deal with
the negative part, we use the generalized BS principle of
Proposition~\ref{BS}; the idea is to treat the difference between
the operator $\RR^{\kappa}_{\alpha,\Gamma}:= \alpha
R^{i\kappa}_{m,m}$ on $L^2(\R)$ and its counterpart corresponding
to $\Gamma_0$ as a perturbation. The integral kernel of the
operator is
$$ \RR^{\kappa}_{\alpha,\Gamma}(s,s') = {\alpha\over 2\pi}\, K_0
\left( \kappa|\gamma(s)\!-\! \gamma(s')| \right) \,, $$
where $K_0$ is the Macdonald function; for $\Gamma=\Gamma_0$ one
has to replace $|\gamma(s)\!-\! \gamma(s')|$ by $|s\!-\! s'|$. In
the last named case the operator is of convolution type and using
Fourier transformation it is easy to check that its spectrum is
absolutely continuous covering the interval $[0,\alpha/2\kappa]$
in correspondence with (\ref{straightspec}).

The key observation is that the kernel of $\DD_{\kappa}:=
\RR^{\kappa}_{\alpha,\Gamma} - \RR^{\kappa}_{\alpha,\Gamma_0}$
satisfies
\begin{equation} \label{Dkern}
\DD_{\kappa}(s,s') := {\alpha\over 2\pi}\, \bigg( K_0 \left(
\kappa|\gamma(s)\!-\! \gamma(s')| \right) - K_0 \left(
\kappa|s\!-\!s'| \right) \bigg) \ge 0 \end{equation}
in view of (a1) and the monotonicity of $K_0$, the inequality
being strict for at least some values of the variables $s,s'$. We
shall then argue in three steps:
\begin{enumerate}
\item a variational argument in combination with (\ref{Dkern})
shows that the spectrum is ``pushed up'' by the perturbation,
$\sup \sigma\left( \RR^{\kappa}_{\alpha,\Gamma} \right)
> {\alpha\over 2\kappa}\,$ if $\,\Gamma$ is not straight.
\item in view of (a2), $\,\DD_{\kappa}\,$ is Hilbert-Schmidt for
$\mu>\,{1\over 2}\,$, and therefore compact.
\item the map $\,\kappa\mapsto \RR^{\kappa}_{\alpha, \Gamma}\,$ is
operator-norm continuous and $\,\RR^{\kappa}_{\alpha,\gamma}\to 0
$ as $\kappa\to\infty$.
\end{enumerate}
The compactness of $\DD_{\kappa}$ implies, in particular, in
combination with Proposition~\ref{BS} the claim about the negative
part of the essential spectrum.

The discrete spectrum part can be also derived from the
generalized BS principle. It follows from the above claims that
there are spectral points of $\RR^{\kappa}_{\alpha,\Gamma}$ above
$\alpha/2\kappa$ and they cannot be anything but eigenvalues of a
finite multiplicity. Moreover, every such eigenvalue depends
continuously on $\kappa$ and tends to zero as $\kappa\to\infty$.
Hence it crosses one at a value $\kappa_0
>\, {1\over 2}\alpha$ giving rise to the sought eigenvalue
of the operator $H_{\alpha,\Gamma}$; for details of the argument
see \cite{EI}.
\end{proof}

It may seem that the result covers only a rather particular class
of graphs. Using the minimax principle, however, we arrive at the
following easy consequence:
\begin{corollary} \label{dsgraph}
Suppose that $\Gamma$ has a subgraph in the form of an infinite
curve satisfying the assumptions of the theorem, and
$\sigma_\mathrm{ess}(H_{\alpha,\Gamma}) = \left\lbrack -\frac14
\alpha^2, \infty \right)$, then the discrete spectrum of
$H_{\alpha,\Gamma}$ is non-empty.
\end{corollary}
It is important that the assumption about preservation of the
essential spectrum can be often verified easily, for instance, in
the situation when $\Gamma$ has outside a compact a finite number
of straight edges separated by non-trivial wedges.

\subsection{An example: leaky star graphs} \label{ss: star}

To illustrate the last made claim, let us investigate in more
detail a particular class of such graphs, namely the situation
when $\Gamma$ is of a \emph{star shape}. Given an integer $N\ge
2$, consider an $(N\!-\!1)$-tuple $\beta=\{ \beta_1,\dots,
\beta_{N-1}\}$ of positive numbers such that
$$ \beta_N := 2\pi - \sum_{j=1}^{N-1} \beta_j > 0\,.  $$
Denote $\vartheta_j:= \sum_{i=1}^j \beta_j$ and $\vartheta_0:=0$.
Let $L_j$ be the radial half-line with the endpoint at the origin,
$L_j:= \{ x\in\R^2:\; \arg x=\vartheta_j \}$, naturally
parametrized by its arc length $s=|x|$. These half-lines will be
the edges of $\Gamma \equiv \Gamma_{\beta} := \bigcup_{j=0}^{N-1}
L_j$ to which we associate\footnote{Properties of $H_N(\beta)$
depend on the order of the angles, however, operators related by a
cyclic permutation are unitarily equivalent by an appropriate
rotation of the plane.} the Hamiltonian
$H_N(\beta):=H_{\alpha,\Gamma_\beta}$. Trivial examples are
\begin{enumerate}
\item $\,H_2(\pi)\,$ corresponding to a straight line which has
obviously a purely a.c. spectrum covering the interval $[-\frac14
\alpha^2, \infty)$,
\item $\,H_4(\beta_s)\,$ with $\beta_s= \left\lbrace {\pi\over 2},
{\pi\over 2}, {\pi\over 2} \right\rbrace$ corresponding to
cross-shaped $\Gamma$ allows again a separation of variables. The
a.c. part of its spectrum is the same as above, and in addition,
there is a single isolated eigenvalue $-\frac12\alpha^2$
corresponding to the eigenfunction $(2\alpha)^{-1}
\e^{-\alpha(|x|+|y|)/2}$.
\end{enumerate}
Star-shaped graphs have the property indicated above:
\begin{proposition} \label{star_ess}
$\,\sigma_\mathrm{ess}(H_N(\beta)) = [-\frac14 \alpha^2, \infty)$
holds for any $N$ and $\beta$.
\end{proposition}
\begin{proof}[Sketch of the proof]
The inequality $\,\inf \sigma_\mathrm{ess}(H_N(\beta)) \ge
-\frac14 \alpha^2\,$ is obtained by Neumann bracketing dissecting
the plane outside a compact into semi-infinite strips with $L_j$
in the middle and ``empty'' wedges. The fact that $[-\frac14
\alpha^2, \infty)$ belongs to the spectrum is checked by means of
a family of Weyl sequences, cf.~\cite{EN1}.
\end{proof}

By Corollary~\ref{dsgraph}, $\:\sigma_\mathrm{disc}
(H_N(\beta))\,$ is nonempty unless $N=2$ and $\beta=\pi$. Using
direct methods one can establish various other properties of the
discrete spectrum\footnote{Variational methods can be also used to
establish the existence of discrete spectrum in some range of
parameters, see the paper \cite{BEW} in this volume.}.
\begin{proposition} \label{star_many_bs}
Fix $N$ and a positive integer $n$. If at least one of the angles
$\beta_j$ is small enough, $\sharp\, \sigma_\mathrm{disc}
(H_N(\beta)) \ge n$. In particular, the number of bound states can
exceed any fixed integer for $N$ large enough.
\end{proposition}
\begin{proof}[Sketch of the proof]
By the minimax principle it is sufficient to check the claim for
$H_2(\beta)$. We choose the coordinate system in such a way that
the two ``arms'' correspond to $\arg\theta = \pm\beta/2$ and
employ trial functions of the form $\Phi(x,y) = f(x)g(y)$
supported in the strip $L\le x \le 2L$, with  $f\in C^2$
satisfying $f(L)=f(2L)=0$, and
$$ g(y) = \left\{ \begin{array}{cll} 1 & \quad \dots \quad &
|y|\le 2d \\ \e^{-\alpha(|y|-2d)} & \quad \dots \quad & |y|\ge 2d
\end{array} \right.
$$
with $d:= L \tan(\beta/2)$. Evaluating the quadratic form of
$H_2(\beta)$ and using minimax principle again we get the result,
see~\cite{EN1} for details.
\end{proof}

  \begin{remark} \label{r:sharp}
The above variational estimate also shows that the bound state
number for a sharply broken line is roughly proportional to the
inverse angle,
$$ n \gtrsim {3^{3/2}\over 8\pi \sqrt{5}}\: \beta^{-1} $$
as $\beta\to 0$. This can be regarded as an expected result, since
the number is given by the length of the effective potential well
which exists in the region where the two lines are so close that
they roughly double the depth of the transverse well.
  \end{remark}

Let us see how the BS equation looks like explicitly for star
graphs. Define
\begin{equation} \label{dist}
d_{ij}(s,s') \equiv d_{ij}^{\beta}(s,s') = \sqrt{s^2\! +{s'}^2\!
-2ss' \cos|\vartheta_j -\vartheta_i|}
\end{equation}
with $\vartheta_j -\vartheta_i = \sum_{l=i+1}^j \beta_l$, in
particular, $d_{ii}(s,s')= |s-s'|$. By $\RR^{\kappa}_{ij}(\beta) =
\RR^{\kappa}_{ji}(\beta)$ we denote the operator $L^2(\R^+) \to
L^2(\R^+)$ with the integral kernel
$$ \RR^{\kappa}_{ij}(s,s';\beta) := {\alpha\over 2\kappa}\: K_0
\left(\kappa d_{ij}(s,s') \right)\,; $$
then the (discrete part of the) spectral problem for the operator
$H_N(\beta)$ is by Proposition~\ref{BS} equivalent to the matrix
integral-operator equation
\begin{equation} \label{speceq1}
\sum_{j=1}^N \left(\RR^{\kappa}_{ij}(\beta) - \delta_{ij}I \right)
\phi_j = 0\,, \quad i=1,\dots,N\,,
\end{equation}
on $\bigoplus_{j=1}^N L^2(\R^+)$. Notice that the ``entries'' of
the above kernel have a monotonicity property,
$\RR^{\kappa}_{ij}(\beta) > \RR^{\kappa}_{ij}(\beta')$ if
$|\vartheta_j -\vartheta_i| < |\vartheta'_j -\vartheta'_i|$. This
fact has the following easy consequence \cite{EN1}:
\begin{proposition} \label{mono2}
Each isolated eigenvalue $\lambda_n(\beta)$ of $H_2(\beta)$ is an
increasing function of the angle $\beta$ between the two
half-lines in $(0,\pi)$.
\end{proposition}

\subsection{Higher dimensions}

If we restrict ourselves to the physically interesting case of
three-dimensional configuration space, there are two possible ways
how to extend the above results. One refers to the situation when
the interaction is supported by a surface. Here unfortunately only
a particular result is known to the date, which we will mention
in Sec.~\ref{ss: 3strong} below.

Let us thus consider the second possibility when $\Gamma$ is an
infinite piecewise $C^1$ curve in $\R^3$. The argument is similar
to that of the previous section but it needs care due to the more
singular character of the interaction. If $\Gamma=\Gamma_0$ the
spectrum of $H_{\alpha,\Gamma}$ is found by separation of
variables; the known properties of the two-dimensional point
interaction \cite{AGHH} imply that
 $$ 
 \sigma (H_{\alpha,\Gamma})=\sigma_\mathrm{ac}
 (H_{\alpha,\Gamma})=[\zeta_{\alpha},\infty )\,,
 $$ 
where $\zeta_{\alpha}=-4 \e^{2(-2 \pi \alpha +\psi (1))}$; in this
expression $-\psi(1)\approx 0.577$ is Euler-Mascheroni constant.
To rephrase it in terms of the BS operator $\QQ_0^\kappa:=
Q_0^{i\kappa}$ notice that the latter equals $\TT_\kappa +\ln
2+\psi(1)$ where $\TT_\kappa$ in the momentum representation acts
as multiplication\footnote{Comparing to
$(2\pi)^{1/2}(p^2+\kappa^2)^{-1/2}$ in the codimension one case.
This is why the kernel of $\QQ_0^\kappa$ makes sense as a
distribution only and a renormalization mentioned in
Remark~\ref{r:3BS} is needed.} by $\ln(p^2+\kappa^2)^{1/2}$.
Consequently, the spectrum of $\QQ_0^\kappa$ is purely absolutely
continuous and equal to $(-\infty ,s_{\kappa}]$ where $s_{\kappa
}:=\frac{1}{2\pi}(\psi(1)-\ln(\kappa/2))$.

To proceed we need assumptions about the curve $\Gamma$. We retain
(a1) and (a3) from the previous section, while (a2) will be
replaced by
\begin{description}
 \item[{\em (a2') ASLS}] there are positive
 $d,\, \mu>\frac12$, and $\omega\in(0,1)$ such that
 $$ 
1-{|\gamma(s)-\gamma(s')|\over|s-s'|} \le d\:\frac{|s\!-\!s'|}
{(1+|s\!-\!s'|) (1+(s^{2}\!+\!s^{\prime 2})^{\mu})^{1/2}}
 $$ 
 holds true in the sector $S_\omega$, the same as before.
\end{description}
In addition to the asymptotic straightness\footnote{For $C^2$
smooth curves we have a sufficient condition for the appropriate
large-distance behavior analogous to that mentioned in the
footnote to assumption (a2).} we require newly some local
smooth\-ness of the curve. Now we can follow proof of
Theorem~\ref{dsexist} step by step; after checking that the
essential spectrum is preserved, and denoting $\QQ^\kappa:=
Q^{i\kappa}$, we prove that
\begin{enumerate}
\item $\sup \sigma (\QQ^{\kappa})>s_{\kappa}$ by a variational
argument using the sign definitess,
 $$ 
 \DD_{\kappa }(s,s^{\prime})= G_{i\kappa}(\gamma
 (s)\!-\!\gamma (s^{\prime}) )- G_{i\kappa }(
 s\!-\!s^{\prime })\ge 0\,,
 $$ 
with a sharp inequlity at least for some values of the variables
$s,s'$.
\item in view of (a2') $\,\DD_{\kappa}\,$ is Hilbert-Schmidt for
$\mu>\,{1\over 2}\,$, hence compact, and the corresponding norm $
\left\| \DD_{\kappa}\right\|_\mathrm{HS}$ is uniformly bounded
w.r.t. $\kappa \geq \left| \zeta_{\alpha}\right|^{1/2}$.
\item the function $\,\kappa \to \QQ^{\kappa}\,$ is operator-norm
continuous in $(\left| \zeta_{\alpha}\right|^{1/2}, \infty)$ and
$\,\QQ^{\kappa}\to -\infty $ holds as $\kappa\to\infty$.
\end{enumerate}
Working out details of this scheme \cite{EK2} we arrive at the
following conclusion:
\begin{theorem} \label{ds3exist}
Fix $\alpha>0$ and suppose that the generating  function $\gamma:
\R\to\R^3$ of $\Gamma$ satisfies the stated assumptions. Then
$\sigma_\mathrm{ess}(H_{\alpha,\Gamma}) = \left\lbrack
\zeta_{\alpha}, \infty \right)$, and the operator
$H_{\alpha,\Gamma}$ has at least one isolated eigenvalue in the
interval $(-\infty ,\zeta_{\alpha})$.
\end{theorem}
Let us remark that the strengthened hypothesis in (a2') is not
needed to prove the existence of the geometrically induced
spectrum, but rather to determine its character; due to the more
strongly singular character of the interaction in the codimension
two case the local smoothness is required to guarantee its
discreteness.

\subsection{Geometric perturbations} \label{ss: geom_pert}

Let us turn to another way in which the leaky character of our
graphs is manifested. Consider a graph $\Gamma\subset\R^\nu$ with
two edges, the endpoints of which are close to each other; we can
think of this situation as of a single edge having a hiatus. In
the standard quantum graph setting, it is only the topology which
matters, either the two edges are connected or not. Here, in
contrast, the distance of the two endpoints plays a role.

Let us first analyze the general codimension one situation, the
$\nu$--dimensional Schr{\"o}dinger operators with a
$\delta$-interaction supported by a $(\nu\!-\!1)$--dimensional
smooth surface having a ``puncture''. We are going to show,
formally speaking, that up to an error term the eigenvalue shift
resulting from removing an $\epsilon$--neighborhood of a surface
point is the same as that of adding a repulsive $\delta$
interaction at this point, with the coupling constant proportional
to the puncture ``area''.

Let $\Gamma$ be a compact $C^{r}$-smooth surface in
$\mathbb{R}^{\nu}$ with $r\ge \frac12\nu$; without loss of
generality we may suppose that it contains the origin. Let further
$\{\PP_{\epsilon}\}_{\epsilon\ge 0}$ be a family of subsets of
$\Gamma$ which obeys the following requirements:
\begin{description}
 \item[{\em (p1) measurability}] $\,\PP_{\epsilon}$ is measurable
 with respect to the $(\nu\!-\!1)$--dimensio\-nal Lebesgue measure
 on $\Gamma$ for any $\epsilon$ small enough.
 \item[{\em (p2) shrinking}] $\,\sup_{x\in \PP_{\epsilon}}|x|=
\OO(\epsilon)$ as $\,\epsilon\to 0$.
\end{description}
Consider the operators $H_{\alpha,\Gamma}$ and $H_{\alpha,
\Gamma_\epsilon}$ corresponding to $\Gamma_\epsilon:= \Gamma
\setminus\PP_\epsilon$ defined as in Sec.~\ref{ss: qf&bc}. Since
$\Gamma_\epsilon$ is bounded, we have
 $$
 \sigma_\mathrm{ess}(H_{\alpha, \Gamma_\epsilon})=[0,\infty)
 \quad \mathrm{and}\quad \sharp\,
 \sigma_\mathrm{disc}(H_{\alpha, \Gamma})<\infty\,. $$
By the minimax principle there is a unique $\alpha^{*}\geq 0$ such
that $\sigma_\mathrm{disc}(H_{\alpha, \Gamma})$ is non-empty if
$\alpha>\alpha^{*}$ while the reverse is true for
$\alpha\leq\alpha^{*}$; it is not difficult to check that
$\alpha^{*}=0$ when $\nu=2$ and $\alpha^{*}>0$ for $\nu\geq 3$,
see \cite[Thm 4.2]{BEKS}.

Let $N$ be the number of negative eigenvalues of $H_{\alpha,
\Gamma}$. Using the convergence of the corresponding quadratic
forms (\ref{Hamform}) on $W^{1,2}(\R^\nu)$ as $\epsilon\to 0$ we
find by \cite[Thm~VIII.3.15]{Ka} that for all $\epsilon$ small
enough $H_{\alpha, \Gamma_\epsilon}$ has the same number $N$ of
negative eigenvalues, which we denote as $\lambda_{1}(\epsilon)<
\lambda_{2}(\epsilon) \leq\cdots\leq\lambda_{N}(\epsilon)$, and
moreover
$$\lambda_{j}(\epsilon)\to\lambda_{j}(0)\quad \mathrm{as}
\quad\epsilon\to 0\quad\mathrm{for}\quad 1\leq j\leq N\,. $$
Let $\{\varphi_{j} (x)\}^{N}_{j=1}$ be an orthonormal system of
eigenfunctions of $H_{\alpha, \Gamma}$ corresponding to these
eigenvalues; without loss of generality we may suppose that
$\varphi_{1}(x)>0\,$ in $\mathbb{R}^{ \nu}$. Using the Sobolev
trace theorem, one can check that each function $\varphi_{j}$ is
continuous on a $\Gamma$--neighborhood of the origin. Given
$\mu\in\sigma_\mathrm{disc}(H_{\alpha, \Gamma})$ we denote as
$m(\mu)$ and $n(\nu)$ the smallest and largest index value $j$,
respectively, for which $\mu=\lambda_{j}(0)$, and we introduce the
positive matrix
 \begin{equation}
C(\mu) := \left(\,\overline{\varphi_{i}(0)}
\varphi_{j}(0)\,\right)_{m(\mu)\leq i,j\leq n(\mu)}\,.
 \end{equation}
denoting by $s_{m(\mu)}\leq s_{m(\mu)+1}\leq\cdots\leq s_{n(\mu)}$
its eigenvalues. In particular, if $\mu=\lambda_j(0)$ is a simple
eigenvalue of $H_{\alpha, \Gamma}$, we have $m(\mu)=n(\mu)=j$ and
$s_j= |\varphi_{j}(0)|^2$. With these prerequisites, we can make
the following claim:
\begin{theorem} \label{surf_hole}
Assume (p1), (p2), and suppose that $\alpha>\alpha^{*}$. For a
given $\mu\in \sigma_\mathrm{disc}(H_{\alpha, \Gamma})$ we have
the asymptotic formula
$$ \lambda_{j}(\epsilon)=\mu +\alpha\,
m_{\Gamma}(\PP_{\epsilon}) s_{j} +o(\epsilon^{\nu-1})\,,\;
m(\mu)\leq j\leq n(\mu)\,, \;\quad\mathit{as}\quad \epsilon\to
0\,,
$$
where $m_{\Gamma}(\cdot)$ stands for the $(\nu\!-\!1)$-dimensional
Lebesgue measure on $\Gamma$.
\end{theorem}
To derive the expansion, one has to note that due to the singular
character of the perturbation a direct use of the asymptotic
perturbation theory of quadratic forms \cite[Sec.~VIII.4]{Ka} is
not possible. Indeed, we have
$$\EE_{-\alpha m_{\Gamma_\epsilon}}[\psi]= \EE_{-\alpha m_{\Gamma}}[\psi]
+\alpha\, m_{\Gamma}(\PP_{\epsilon}) |\psi(0)|^{2}+\OO
(\epsilon^{\nu}) \quad\mathrm{as}\quad\epsilon\to 0
$$
for $\psi\in C^{\infty}_{0}(\mathbb{R}^{\nu})$ and the quadratic
form $C^{\infty}_{0}(\mathbb{R}^{\nu}) \owns \psi\mapsto
|\psi(0)|^{2}\in \mathbb{R}$ does not extend to a bounded form on
$W^{1,2}(\mathbb{R}^{\nu})$, because the set of $\psi\in
C^{\infty}_{0}(\mathbb{R}^{\nu})$ vanishing at the origin is dense
in $W^{1,2}(\mathbb{R}^{\nu})$. A way to eliminate this difficulty
is to employ the compactness of the map $W^{1,2}(\R^{\nu})\owns
f\mapsto f|_{\Gamma}\in L^{2}(\Gamma)$; we refer to \cite{EY4} for
a detailed description of such a proof\footnote{There are other
ways to derive such asymptotic expansions, for instance, the
technique of matching of asymptotic expansions \cite{Il}. The
advantage of the sketched approach is that it requires no
self-similarity properties for the family of shrinking sets
$\PP_\epsilon$.}.

Let us observe further that while the compactness of $\Gamma$ was
used in formulation of the theorem, it played essentially no role
in the proof. This allows us to treat other situations, for
instance, eigenvalues corresponding to non-straight curves as
described in Sec.~\ref{ss: nonstr}. Suppose we replace such a
$\Gamma$ by a family of curves with a hiatus, $\Gamma_\epsilon$
given by the same function $\gamma$ where, however, the argument
runs over $\R\setminus(-\epsilon,\epsilon)$. Using the above
introduced notation, we then have the following result:
\begin{corollary} \label{2hiatus}
Suppose that $\Gamma$ satisfies the assumptions of
Theorem~\ref{dsexist}, then the eigenvalues of $H_{\alpha,
\Gamma}$ and $H_{\alpha, \Gamma_\epsilon}$ obey the asymptotic
formula
$$ \lambda_{j}(\epsilon)=\lambda_{j}(0) +2\alpha\epsilon\,
s_j+ o(\epsilon) \,,\; m(\mu)\leq j\leq n(\mu)\,,
\;\quad\mathit{as}\quad \epsilon\to 0
$$
\end{corollary}

 \begin{remark} \label{3hiatus}
As before, things look differently in the case of codimension two.
If we have, for instance, a simple eigenvalue of $H_{\alpha,
\Gamma}$ with the eigenfunction $\varphi$, where $\Gamma$ is a
curve in $\R^3$ and perturb the latter by making a
$2\epsilon$--hiatus in it, the leading term in the perturbation
expansion is again proportional to $|\varphi(0)|^2$, however, this
time it comes multiplied not by $\epsilon$ but rather
$\epsilon\ln\epsilon$ -- cf.~\cite{EK4}.
 \end{remark}

\subsection{An isoperimetric problem} \label{ss: isoper}

The above results do not exhaust ways in which the edge shapes
influence spectral properties of leaky-graph Hamiltonians. Let us
mention one more; for simplicity, we restrict ourselves again to
the planar case, $\Gamma\subset\R^2$. If $\Gamma$ is of a finite
length, we have $\sigma_\mathrm{ess}(H_{\alpha,\Gamma})=
[0,\infty)$, while the discrete spectrum is nonempty and finite,
so that
 \begin{equation} \label{ex_ground}
 \lambda_1 \equiv \lambda_1(\alpha,\Gamma):= \inf \sigma
 \left(H_{\alpha,\Gamma}\right)<0\,.
 \end{equation}
Suppose now that $\Gamma$ is a loop of a fixed length and ask
which shape of it makes the above principal eigenvalue maximal.

Let us make the assumptions more precise. We suppose that
$\gamma:\: [0,L]\to \R^2$ is a closed $C^1$, piecewise $C^2$
smooth curve, $\gamma(0)= \gamma(L)$; we allow self-intersections
provided the curve meets itself at a non-zero angle. Furthermore,
we introduce the equivalence relation: the loops $\Gamma$ and
$\Gamma'$ belong to the same class if one can be obtained from the
other by a Euclidean transformation of the plane. Spectral
properties of the corresponding operators $H_{\alpha,\Gamma}$ and
$H_{\alpha,\Gamma'}$ are obviously the same, hence we will speak
about a curve $\Gamma$ having in mind the corresponding
equivalence class. The stated assumptions are satisfied, in
particular, by the circle, say $\CC:= \{\,({L\over 2\pi}\cos
\frac{2\pi s}{L}, {L\over 2\pi}\sin \frac{2\pi s}{L}):\:
s\in[0,L]\,\}$, and its equivalence class.
 \begin{theorem} \label{isoper}
 Within the above described class of loops, the principal eigenvalue
 $\lambda_1(\alpha,\Gamma)$ is for any fixed $\alpha>0$ and $L>0$
 sharply maximized by the circle.
 \end{theorem}
We will need the following geometric result about means of chords:
 \begin{proposition} \label{chord}
Let $\Gamma$ have the properties described above, then
  $$
 \int_0^L |\gamma(s\!+\!u) -\gamma(s)|^p\,
 \D s \,\le\, \frac{L^{1+p}}{\pi^p} \sin^p \frac{\pi u}{L}\quad
 \mathrm{for} \quad p\in(0,2]\,.
  $$
 \end{proposition}
The right-hand side of the inequality is obviously the value of
the integral for the circle. Notice that the same is true for
loops in $\R^\nu$ and a similar \emph{reverse} inequality holds
for negative powers $p\in[-2,0)$. To prove this it is only
necessary to establish the result for $p=2$ which was done in
various ways in \cite{Lu, ACF, EHL}, see also \cite{Ex2} for a
local maximum proof.

\begin{proof}[Sketched proof of Theorem~\ref{isoper}]
We shall rely again on Proposition~\ref{BS}, which relates our
eigenvalue problem to the integral equation
$\RR_{\alpha,\Gamma}^\kappa \phi= \phi$ on $L^2([0,L])$ with
$\RR_{\alpha,\Gamma}^\kappa$ defined similarly as in the proof of
Theorem~\ref{dsexist}; we note that the operator-valued function
$\kappa\mapsto \RR_{\alpha,\Gamma}^\kappa$ is strictly decreasing
in $(0,\infty)$ and $\|\RR_{\alpha,\Gamma}^\kappa\|\to 0$ as
$\kappa\to\infty$. By a positivity improving argument the maximum
eigenvalue of $\RR_{\alpha,\Gamma}^\kappa$ is simple, and the same
is true by Proposition~\ref{BS} for the ground state of
$H_{\alpha,\Gamma}$. If $\Gamma$ is a circle, the latter exhibits
rotational symmetry, and using Proposition~\ref{BS} again we see
that the respective eigenfunction of
$\RR_{\alpha,\CC}^{\tilde\kappa_1}$ corresponding to the unit
eigenvalue is constant, $\tilde \phi_1(s)= L^{-1/2}$. Then we have
 $$ 
 \max \sigma(\RR_{\alpha,\CC}^{\tilde\kappa_1}) = (\tilde\phi_1,
 \RR_{\alpha,\CC}^{\tilde\kappa_1} \tilde\phi_1) = \frac1L
 \int_0^L \int_0^L
 \RR_{\alpha,\CC}^{\tilde\kappa_1}(s,s') \,\D s\D s'\,,
 $$ 
while for a general $\Gamma$ a simple variational estimate gives
 $$ 
 \max \sigma(\RR_{\alpha,\Gamma}^{\tilde\kappa_1}) \ge (\tilde\phi_1,
 \RR_{\alpha,\Gamma}^{\tilde\kappa_1} \tilde\phi_1) = \frac1L
 \int_0^L \int_0^L
 \RR_{\alpha,\Gamma}^{\tilde\kappa_1}(s,s') \,\D s\D s'\,;
 $$ 
hence to check that the circle is a maximizer it sufficient to
show that
 $$ 
 \int_0^L \int_0^L
 K_0\big(\kappa|\Gamma(s) \!-\!\Gamma(s')|\big) \,\D s\D s' \ge
 \int_0^L \int_0^L
 K_0\big(\kappa|\CC(s) \!-\!\CC(s')|\big) \,\D s\D s'
 $$ 
holds \emph{for all} $\kappa>0$ and $\Gamma$ of the considered
class. By a simple change of variables we find that this is
equivalent to positivity of the functional
 $$ 
 F_\kappa(\Gamma):= \int_0^{L/2} \D u \int_0^L \D s \bigg[
 K_0\big(\kappa|\Gamma(s\!+\!u) -\Gamma(s)|\big) -
 K_0\big(\kappa|\CC(s\!+\!u) -\CC(s)|\big) \bigg]\,,
 $$ 
where the second term is equal to $K_0\big(\frac{\kappa L}{\pi}
\sin \frac{\pi u}{L}\big)$. Now we employ the (strict) convexity
of $K_0$ which yields by means of the Jensen inequality the
estimate
 $$ 
 \frac 1L \,F_\kappa(\Gamma)\ge \int_0^{L/2} \left[
 K_0\left( \frac{\kappa}{L} \int_0^L |\Gamma(s\!+\!u) -\Gamma(s)| \D s
 \right) - K_0\left(\frac{\kappa L}{\pi} \sin \frac{\pi
 u}{L}\right) \right]\, \D u\,,
 $$ 
where the inequality is sharp unless $|\Gamma(s\!+\!u) -\Gamma(s)|
\D s$ is independent of~$s$. Finally, we note that $K_0$ is
decreasing in $(0,\infty)$, hence the result follows from the
geometric inequality of Proposition~\ref{chord} with $p=1$, see
\cite{Ex2} for details.
\end{proof}

\subsection{Scattering} \label{ss: scatt}

The investigation of leaky graphs is not exhausted, of course, by
analysis of their discrete spectrum. Another important problem
concerns scattering on graphs having semi-infinite edges. Our
knowledge about this subject is far from satisfactory at present
and we will concentrate here on a particular situation. First of
all, we consider again planar graphs, $\Gamma\subset\R^2$, only.
Secondly, we restrict ourself to graphs which can be regarded as a
local modification of a straight line. And finally, we will
analyze the situation which is from the point of view of our
physical motivation the most interesting, namely the negative part
of the spectrum where the scattering states are ``guided'' along
the graph edges.

The unperturbed graph is thus the straight line $\Sigma =\{(x_1
,0):\,x_{1}\in \R\}$ for which $H_{\alpha,\Sigma}$ allows a
separation of variables; the spectrum is purely a.c. and, in
particular, the generalized eigenfunctions corresponding to
$\lambda \in (-\frac14\alpha^2, 0)$ are
\begin{equation}\label{gensigma}
\omega_{\lambda}(x_1 ,x_2 )=\mathrm{e}^{i(\lambda +\alpha^2 /4
)^{1/2}x_{1}}\mathrm{e}^{-\alpha|x_{2}|/2}
\end{equation}
and its complex conjugate $\bar\omega_{\lambda}$. If we perturb
the line $\Sigma$ locally, we will get a nontrivial scattering,
but the essentially one-dimensional character of the motion will
remain asymptotically preserved. Let us specify the perturbation:
\begin{description}
 \item[{\em (s1) locality}] there is a compact $M\subset\R^2$
 such that $\Gamma \setminus M= \Sigma \setminus M$.
 \item[{\em (s2) finiteness}] $\,\Gamma \setminus \Sigma$ is a
 finite graph with the properties (g1) and (g2).
\end{description}
We will treat the operator $H_{\alpha,\Gamma}$ defined by the
prescription given in Sec.~\ref{ss: qf&bc} as a singular
perturbation of $H_{\alpha,\Sigma}$ supported by the set
\begin{equation}\label{Lambda}
\Lambda =\Lambda_0 \cup \Lambda _1 \quad \mathrm{with}\quad
\Lambda_{0}:=\Sigma \setminus \Gamma \,,\,\,\, \Lambda_{1}:=\Gamma
  \setminus \Sigma = \bigcup_{i=1}^{N}\Gamma _{i}\,;
\end{equation}
the coupling constant of the perturbation will take the positive
value $\alpha$ on the ``erased'' part $\Lambda_{0}$ and negative
one, $-\alpha$, on added edges $\Lambda_{1}$.

For our present purpose we need a more suitable resolvent
expression than that given by Proposition~\ref{BS}; instead of
(\ref{freeG}) we will use the resolvent of $H_{\alpha,\Sigma}$ as
the comparison operator. The latter can be expressed, of course,
again by Proposition~\ref{BS}: we have $R_{\Sigma}^{k}
=R_0^{k}+\alpha R^{k}_{\mu}(I-\alpha R^{k}_{\mu,\mu} )^{-1}
(R^{k}_{\mu})^{*}$, where for simplicity we write $R^{k}_{\D
x,\mu}\equiv R^{k}_{\mu}$ and $\mu:=m_\Sigma$ is the measure
associated with the line by (\ref{meas}). A direct calculation
then yields the expression
\begin{equation}\label{kernelsig}
R^{k}_{\Sigma}(x\!-\!y)=G_{k}(x\!-\!y)+\frac{\alpha}{4\pi^{3}}
\int_{\R^{3}}\frac{\mathrm{e}^{ipx-ip'y}}{(p^{2}\!-\!k^{2})
(p'^{2}\!-\!k^{2})} \frac{\tau _{k}(p_{1})}{2\tau
_{k}(p_{1})\!-\!\alpha}\, \mathrm{d}p\, \mathrm{d}p'_{2}
\end{equation}
with $\tau _{k}(p_{1}) :=(p^{2}_{1}-k^{2}) ^{1/2}$ for any $k$
with $\mathrm{Im}\,k>0$ such that $k^2\in \C\setminus
[-\frac14\alpha^2 ,\infty )$. To get the above indicated
expression for the resolvent of $H_{\alpha,\Gamma}$ we decompose
the Dirac measure associated with the perturbation by (\ref{meas})
as
\begin{equation}\label{decompnu}
\nu \equiv \nu_\Lambda =\nu_{0}+\sum_{i=1}^{N}\nu_{i}\,,
\end{equation}
where $\nu_0$ corresponds to the ``erased'' edge $\Lambda_0$ and
$\nu_i$ to $\Gamma_i$. The associated integral operator will act
in the Hilbert space $\mathrm{h}:= L^{2}(\nu)$ which inherits from
(\ref{decompnu}) the decomposition $\mathrm{h}=\mathrm{h}_{0}
\oplus \mathrm{h}_{1}$ with $\mathrm{h}_{0}:= L^{2}(\nu_{0})$ and
$\mathrm{h}_{1}:= \bigoplus_{i=1}^{N}L^{2}(\nu_{i})$. We will
again need the trace maps, this time of the operator associated
with (\ref{kernelsig}), given by
 \begin{equation} \label{tracop}
R_{\Sigma ,\nu}^{k}:\mathrm{h}\rightarrow L^2 \,,\quad
R_{\Sigma,\nu}^{k}f =R^{k}_{\Sigma}\ast f\nu\quad
\mathrm{for}\quad f\in \mathrm{h}
 \end{equation}
together with the adjoint $(R_{\Sigma ,\nu}^{k})^{\ast}: L^2
\rightarrow \mathrm{h}$ and $R_{\Sigma ,\nu \nu}^{k}$ which is the
operator-valued matrix in $\mathrm{h}$ with the ``block elements''
$R^{k}_{\Sigma ,ij}: L^{2}(\nu_{j})\to L^{2}(\nu_{i})$ defined as
the appropriated embeddings of (\ref{kernelsig}). They have
properties analogous to those of Proposition~\ref{BS} which can be
checked in a similar way, cf.~\cite{EK5}.

\begin{proposition} \label{proposi1}
The operator $R_{\Sigma ,\nu}^{i\kappa}$ is bounded for any
$\kappa \in (\frac12\alpha, \infty)$. Moreover, to any $\sigma
>0$ there is a $\kappa_\sigma >0$ such that $\|R_{\Sigma ,\nu
\nu}^{i\kappa}\|<\sigma$ holds for $\kappa > \kappa_\sigma$.
\end{proposition}

To express the resolvent we introduce an operator-valued matrix in
$\mathrm{h}=\mathrm{h}_{0}\oplus\mathrm{h}_{1}$,
\begin{equation} \label{auxmatrix}
\Theta_\alpha^{k} :=-(\alpha ^{-1}\mathbb{I}
+R^{k}_{\Sigma,\nu\nu})  \quad \mathrm{with} \quad \mathbb{I}:=
\left(
\begin{array}{cc}I_{0}
& 0   \\
0 & -I_{1}  \end{array} \right)\,,
\end{equation}
where $I_{i}$ are the unit operators in $\mathrm{h}_{i}$. By
Proposition~\ref{proposi1}, the operator $\Theta_\alpha^{i\kappa}$
is boundedly invertible for $\kappa $ large enough and we have the
following theorem \cite{EK5}:

\begin{theorem}\label{res_scatt}
Suppose that $(\Theta_\alpha^{k})^{-1}\in \mathcal{B}(\mathrm{h})$
hold for $k\in \C^+$ and the operator
 $$ 
  R^{k}_{\Gamma }=R^{k}_{\Sigma }+ R^{k}_{\Sigma ,\nu}
  (\Theta_\alpha^{k})^{-1}(R^{k}_{\Sigma ,\nu})^{*}
 $$ 
is defined everywhere in $L^2(\R^2)$. Then $k^{2}$ belongs to
$\rho (H_{\alpha,\Gamma})$ and the resolvent $(H_{\alpha,
\Gamma}-k^2)^{-1}$ coincides with $R^{k}_{\Gamma }$.
\end{theorem}

A simple estimate shows that the operator $R^{k}_{\Sigma,\nu}$ is
Hilbert--Schmidt under our assumptions and since the other two
factors are bounded by Proposition~\ref{proposi1}, we get as a
consequence stability of the essential spectrum.
\begin{corollary} \label{essspec}
$ \sigma _{\mathrm{ess}}(H_{\alpha,\Gamma}) =\sigma
_{\mathrm{ess}}(H_{\alpha,\Sigma})=\left[-\frac{1}{4}\alpha ^2,
\infty \right)\,. $
\end{corollary}

Let us turn now to the proper topic of this section, which is the
scattering theory for the pair $(H_{\alpha,\Gamma},
H_{\alpha,\Sigma})$. To establish existence of the wave operators
by the standard Birman-Kuroda method we need to check that the
resolvent difference $B^{k}:= R^{k}_{\Sigma,\nu}
(\Theta_\alpha^{k})^{-1} (R^{k}_{\Sigma,\nu})^{*}$ is of the trace
class.
\begin{proposition} \label{trcl}
$B^{i\kappa}$ is a trace class operator for all $\kappa$
sufficiently large.
\end{proposition}
\begin{proof}[Sketch of the proof]
The idea is inspired by \cite{BT}. We estimate the
(sign-indefinite) operators $(\Theta_\alpha^{i\kappa})^{-1}$ and
$B^{i\kappa}$ from above and below by a positive and negative
operator, respectively, which are obtained by taking both signs in
the matrix multiplying $\alpha^{-1}$ in (\ref{auxmatrix}) the
same. Then we can integrate the kernel diagonal of the estimates
to $B^{i\kappa}$ with a suitable cut-off, using the lemma
following Thm~XI.31 in \cite{RSi}, and to show subsequently that
the trace-class property is preserved when the cut-off is removed,
cf.~\cite{EK5} for details.
\end{proof}

The existence of wave operators which follows from
Proposition~\ref{trcl} does not tell us much, and we have to find
also the on--shell S-matrix relating the incoming and outgoing
asymptotic solutions. In particular, for scattering in the
negative part of the spectrum with a fixed $\lambda \in
(-\frac{1}{4}\alpha ^2,0)$ corresponding to the effective momentum
$k_{\alpha }(\lambda ):=(\lambda +\alpha^{2}/4)^{1/2}$, the latter
are combinations of $\omega_{\lambda}$ and $\bar\omega_{\lambda}$
given by (\ref{gensigma}). These generalized eigenfunctions and
their analogues $\omega_z$ for complex values of the energy
parameter are $L^2$ only locally, of course, but we can use the
standard trick and approximate them by regularized functions, for
instance
$$
\omega _{z}^{\delta }(x)=\mathrm{e}^{-\delta
x_{1}^{2}}\omega_{z}(x)\quad \mathrm{with} \quad z\in \rho
(H_{\alpha,\Sigma})\,,
$$
which naturally belong to the domain $D(H_{\alpha,\Sigma})$. Now
we are looking for a function $\psi _{z}^{\delta }$ such that
$(-\Delta _{\Gamma }-z)\psi _{z}^{\delta }=(-\Delta _{\Sigma
}-z)\omega _{z}^{\delta }$. Computing the right-hand side and
taking the limit $\lim_{\epsilon \to 0}\psi_{\lambda +i\epsilon
}^{\delta }=:\psi_{\lambda }^{\delta }$ in the topology of $L^2$
we find that $\psi_{\lambda }^{\delta }$ still belongs to
$D(H_{\alpha,\Gamma})$, and moreover
$$
\psi_{\lambda }^{\delta }=\omega _{\lambda }^{\delta }+ R_{\Sigma
,\nu}^{k_\alpha(\lambda )} (\Theta^{k_\alpha(\lambda
)})^{-1}I_{\Lambda} \omega_{\lambda }^{\delta }\,,
$$
where $I_{\Gamma}$ is the standard embedding from $W^{1,2}$ to
$\mathrm{h}=L^{2}(\nu_{\Lambda})$ and $R_{\Sigma,\nu}^{
k_\alpha(\lambda )}$ is the integral operator acting on the
Hilbert space $\mathrm{h}$, analogous to (\ref{tracop}), with the
kernel
\begin{equation}\label{limitspec}
R^{k_\alpha(\lambda )}_{\Sigma }(x\!-\!y):=\lim_{\epsilon \to 0}
R^{k_\alpha(\lambda +i\epsilon)}_{\Sigma}(x\!-\!y)\,;
\end{equation}
similarly $\Theta^{k_\alpha(\lambda )}:= -\alpha ^{-1} \mathbb{I}
-R^{k_\alpha(\lambda )}_{\Sigma ,\nu\nu}$ are the operators on
$\mathrm{h}$ with $\mathrm{R}^{k_\alpha(\lambda )}_{\Sigma
,\nu\nu}$ being the embeddings defined by means of
(\ref{limitspec}). When we remove the regularization, the
pointwise limit $\psi _{\lambda } :=\lim_{\delta\to 0 }\psi
_{\lambda }^{\delta}$ ceases to be square integrable, however, it
still belongs locally to $L^2$ and yields the generalized
eigenfunction of $H_{\alpha,\Gamma}$, namely
\begin{equation}\label{geneg2}
\psi_{\lambda }=\omega_{\lambda }+ R_{\Sigma,\nu}^{
k_\alpha(\lambda )} (\Theta^{k_\alpha(\lambda )})^{-1}J_{\Lambda}
\omega_{\lambda }\,,
\end{equation}
where $J_{\Lambda} \omega_{\lambda }$ is the embedding of
$\omega_{\lambda }$ to $L^{2}(\nu_{\Lambda})$. The on--shell
S-matrix can be then found by inspecting the asymptotic behavior
of the function $\psi _{\lambda }$ as $|x_{1}|\to \infty$. Using
the explicit form of the kernel (\ref{limitspec}) derived in
\cite{EK2} one arrives by a direct computation at the following
result:

\begin{theorem} \label{geneigenvTH}
For a fixed $\lambda \in (-\frac{1}{4}\alpha^2,0)$ the generalized
eigenfunctions of $H_{\alpha,\Gamma}$ behave under the assumptions
(s1), (s2) asymptotically as
 $$ 
\psi_\lambda(x) \approx \left\{ \begin{array}{lcl}
\mathcal{T}(\lambda )\,\mathrm{e}^{ik_{\alpha }(\lambda)x_{1}} \,
\mathrm{e}^{-\alpha|x_{2}|/2} & \;\; \mathrm{for } \;\; & x_{1}\to
+\infty \\ [.3em]
\mathrm{e}^{ik_{\alpha}(\lambda)x_{1}}\mathrm{e}^{-\alpha|x_{2}|/2}
+ \mathcal{R}(\lambda )\,\mathrm{e}^{-ik_{\alpha
}(\lambda)x_{1}}\mathrm{e}^{-\alpha|x_{2}|/2} & \;\; \mathrm{for
}\;\; & x_{1}\to -\infty \end{array} \right.
 $$ 
where $k_{\alpha}(\lambda):=(\lambda +\alpha ^{2}/4)^{1/2}$ is the
effective momentum along $\Sigma$ and $\mathcal{T}(\lambda
)\,,\mathcal{R}(\lambda )$ are the transmission and reflection
amplitudes, respectively, given by
$$
\mathcal{R}(\lambda )= 1-\mathcal{T}(\lambda)=
\frac{i\alpha}{8k_{\alpha}(\lambda)}\,
\left((\Theta^{k_\alpha(\lambda )})^{-1} J_\Lambda \omega
_{\lambda },J_\Lambda \bar{\omega} _{\lambda
}\right)_{\mathrm{h}}\,.
$$
\end{theorem}

\section{Strong-coupling asymptotics} \label{s: strong}

The coupling constant in $H_{\alpha,\Gamma}$ determines how is the
particle attracted to the graph, and this in turn implies, in
particular, which is the ``spread'' of possible eigenfunctions in
the direction transverse to the edges. It is thus natural to ask
what happens in the case of a strong coupling when the wave
functions are transversally sharply localized. We are going now to
show that if the interaction support is a sufficiently smooth
manifold, various asymptotic formul{\ae} can be derived.

\subsection{Interactions supported by curves} \label{ss: strong
curve}

In distinction to the previous considerations it is rather the
dimension than codimension of $\Gamma$ that will be important. As
usual, we begin with the case of planar curves, at first finite
ones.

\begin{theorem} \label{loop2_asympt}
Suppose that $\gamma:\: [0,L]\to\R^2$ is a $C^4$ smooth function,
$|\dot\gamma|=1$, which defines a curve $\Gamma$; then the
relation
$$ \sharp\,\sigma_\mathrm{disc} (H_{\alpha,\Gamma}) =
\frac{\alpha L}{2\pi} +\mathcal{O}(\ln\alpha) $$
holds as $\alpha\to\infty$. In addition, if $\Gamma$ is a closed
curve without self-intersections, then the $j$-th eigenvalue of
the operator $H_{\alpha,\Gamma}$ behaves asymptotically as
$$ \lambda_j(\alpha) = -\frac{1}{4}\,\alpha^2 + \mu_j +
\mathcal{O}(\alpha^{-1} \ln\alpha)\,, $$
where $\mu_j$ is the $j$-th eigenvalue of the operator $S_\Gamma
:= -{\D^2\over \D s^2} - {1\over 4}k(s)^2$ on $L^2(0,L)$ with
periodic b.c., counted with multiplicity, and $k(s)$ is the signed
curvature of $\Gamma$.
\end{theorem}
\begin{proof}[Sketch of the proof]
Suppose first that $\Gamma$ is closed, without
self-intersec\-tions, and consider its strip neighborhood
analogous to (\ref{curvi}), in other words, the set $\Sigma_{a}$
onto which the function $\,\Phi_{a}:\: [0,L)\times (-a,a) \to \R^2
$ defined by
 \begin{equation} \label{strip map}
 (s,u)\mapsto (\gamma_{1}(s)-u\gamma^{\prime}_{2}(s),
 \gamma_{2}(s)+u\gamma^{\prime}_{1}(s))
 \end{equation}
maps, diffeomorphically for all $a>0$ small enough. The main idea
is to apply to $H_{\alpha,\Gamma}$ the Dirichlet-Neumann
bracketing at the boundary of $\Sigma_{a}$,
$$ (-\Delta^{ \mathrm{N}}_{\Lambda_a}) \oplus L_{a,\alpha}^{-}
\leq H_{\alpha,\Gamma} \leq (-\Delta^{\mathrm{D}}_{\Lambda_a})
\oplus L_{a,\alpha}^{+}, $$
where $\Lambda_a= \Lambda^{\mathrm{in}}_{a} \cup
\Lambda^{\mathrm{out}}_{a}$ is the exterior domain, and
$L_{a,\alpha}^{\pm}$ are self-adjoint operators associated with
the forms
$$ q_{a,\alpha}^{\pm}[f] =\Vert\nabla
f\Vert^{2}_{L^{2}(\Sigma_{a})} -\alpha\int_{\Gamma}|f(x)|^{2}
\,\mathrm{d}S $$
where $f\in W_0^{1,2}(\Sigma_a)$ and $W^{1,2}(\Sigma_a)$ for
$\pm$, respectively. The exterior $\R^2\setminus
\overline{\Sigma}_a$ does not contribute to the negative part of
the spectrum, so we may consider $L_{a,\alpha}^{\pm}$ only.

We use the curvilinear coordinates $(s,u)$, the same as in
(\ref{curvi}), passing from $L_{a,\alpha}^{\pm}$ to unitarily
equivalent operators given by quadratic forms
\begin{eqnarray*}  \lefteqn{ b^{+}_{a,\alpha}[f] =
\int^{L}_{0}\int^{a}_{-a}(1+uk(s))^{-2} \left|\frac{\partial
f}{\partial s}\right|^{2}(s,u)\,\mathrm{d}u\,\mathrm{d}s
+\int^{L}_{0}\int^{a}_{-a} \left|\frac{\partial f}{\partial
u}\right|^{2}(s,u)\,\mathrm{d}u\,\mathrm{d}s } \\ &&
+\int^{L}_{0}\int^{a}_{-a}V(s,u)|f(s,u)|^{2}\,\mathrm{d}s\,\mathrm{d}u
-\alpha\int^{L}_{0}|f(s,0)|^{2}\,\mathrm{d}s \phantom{AAAAAAAAAA}
\end{eqnarray*}
with $f\in W^{1,2}((0,L)\times(-a,a))$ satisfying periodic
boundary conditions in the variable $s$ and Dirichlet b.c. at
$u=\pm a$, and
$$  b^{-}_{a,\alpha}[f] =
b^{+}_{a,\alpha}[f] - \sum_{j=0}^1 \frac{1}{2} (-1)^j \int^{L}_{0}
\frac{k(s)}{1+(-1)^jak(s)}\,|f(s,(-1)^ja)|^{2}\,\mathrm{d}s\,,  $$
where $V$ is the usual curvature induced potential \cite{ES}
\begin{equation} \label{effpot2}
V(s,u)= - {k(s)^2\over 4(1\!+\!uk(s))^2} +
{uk^{\prime\prime}(s)\over 2(1\!+\!uk(s))^3} -
{5u^{2}k^{\prime}(s)^{2}\over 4(1\!+\!uk(s))^4}\,.
\end{equation}
We may employ rougher bounds squeezing $H_{\alpha,\Gamma}$ between
$\tilde{H}^{\pm}_{a,\alpha} =U^{\pm}_{a}\otimes 1+1\otimes
T^{\pm}_{a,\alpha}$ with decoupled variables. Here $U^{\pm}_{a}$
are self-adjoint operators on $L^2(0,L)$ given by
$$ U^{\pm}_{a}=-(1\mp a\|k\|_\infty)^{-2} \frac{\D^{2}}
{\D s^{2}}+V_{\pm}(s) $$
with periodic b.c., where $V_-(s)\le V(s,u) \le V_+(s)$ with an
$\mathcal{O}(a)$ error, and the transverse operators are
associated with the forms
$$ t^{+}_{a,\alpha}[f]=
\int^{a}_{-a}|f^{\prime}(u)|^{2}\,\mathrm{d}u-\alpha |f(0)|^{2}
$$
and
$$ t^{-}_{a,\alpha}[f]= t^{-}_{a,\alpha}[f] -
\|k\|_\infty(|f(a)|^2 +|f(-a)|^2)\,,$$
where $f\in W^{1,2}_0(-a,a)$ and $W^{1,2}(-a,a)$ for the $\pm$
sign, respectively. Their negative spectrum can be localized with
an exponential precision: there is a $c>0$ such that
$T_{\alpha,a}^{\pm}$ has for $\alpha$ large enough a single
negative eigenvalue $\kappa_{\alpha,a}^{\pm}$ satisfying
\begin{equation} \label{transest2}
-\frac{\alpha^{2}}{4} \left(1+c\, \mathrm{e}^{-\alpha a/2} \right)
<\kappa_{\alpha,a}^{-} < -\frac{\alpha ^{2}}{4}<
\kappa_{\alpha,a}^{+} < -\frac{\alpha ^{2}}{4} \left(1 -8
\mathrm{e}^{-\alpha a/2} \right)
\end{equation}
To finish the proof, one has to check that the eigenvalues of
$U^{\pm}_a$ differ by $\mathcal{O}(a)$ from those of the
comparison operator, then we choose $a= 6\alpha^{-1} \ln\alpha$ as
the neighborhood width and putting the estimates together we get
the eigenvalue asymptotic formula; for details see \cite{EY1}. If
 $\Gamma$ is not closed, the same can be done with the
comparison operators $S_\Gamma^\mathrm{D,N}$ having the
appropriate b.c., Dirichlet or Neumann, at the endpoints of
$\Gamma$; this gives the estimate on $\sharp\,\sigma_\mathrm{disc}
(H_{\alpha,\Gamma})$.
\end{proof}

The case of a finite curve in $\R^3$ is similar, but we have to be
more cautious about the regularity of the curve. It will be again
a graph of a $C^4$ smooth function, $\gamma:\: [0,L]\to \R^{3}$
with $|\dot\gamma(s)|=1$. To construct the three-dimensional
counterpart of the ``straightening'' transformation used in the
above proof, we suppose for simplicity that $\Gamma$ possesses a
global Frenet frame\footnote{This assumption can be weakened, see
the footnote in Sec.~\ref{ss: cod2}.} and consider the map
$\phi_a: \:[0,L]\times\BB_{a}\to\R^3$
$$
\phi_{a}(s,r,\theta)= \gamma (s)-r \left\lbrack n(s)
\cos(\theta\!-\!\beta(s))+b(s)\sin(\theta\!-\!\beta(s))
\right\rbrack\,,
$$
where $\BB_{a}$ is the circle of radius $a$ centered at the origin
and the function $\beta$ has to be specified; for small enough $a$
it is a diffeomorphic map on a tubular neighborhood $\Sigma_a$ of
$\Gamma$ which does not intersect itself. The geometry of
$\Sigma_a$ is naturally described in terms of the metric tensor
$g_{ij}$ expressed by means of the curvature $k$ and torsion
$\tau$ of $\Gamma$. In particular, in the neighborhood with a
circular cross section we can always choose the so-called Tang
coordinate system, $\dot\beta=\tau$, in which the tensor $g_{ij}$
is diagonal, i.e. the longitudinal and transverse variable
decouple \cite{DE}.

To state the result we have to note that in the codimension two
case the strong coupling means large \emph{negative} values of the
parameter $\alpha$.

\begin{theorem} \label{loop3_asympt}
For curves $\Gamma$ without self-intersections described above, we
have
$$ \sharp\,\sigma_\mathrm{disc}(H_{\alpha,\Gamma})=\frac{L}{\pi
}(-\zeta_{a})^{1/2}(1+\mathcal{O} (e^{\pi \alpha })). $$
as $\alpha\to-\infty$, where we put again $\zeta_{\alpha}:= -4
\e^{2(-2 \pi \alpha +\psi (1))}$. If, in addition, $\Gamma$ is a
closed curve, the $j$-th eigenvalue of the operator
$H_{\alpha,\Gamma}$ behaves asymptotically as
$$ \lambda_{j}(\alpha)=\zeta_{\alpha } +\mu_{j}
+\mathcal{O}(\e^{\pi \alpha})\,, $$
where $\mu_j$ is the $j$-th eigenvalue of the same operator
$S_\Gamma$ as in Theorem~\ref{loop2_asympt}.
\end{theorem}
\begin{proof}[Sketch of the proof]
The argument follows the same scheme. We use Dirichlet--Neumann
bracketing at the boundary of $\Sigma_a$ and estimate the internal
part using the Tang coordinate system. The effective potential
replacing (\ref{effpot2}) is known from \cite{DE}; it is important
that the torsion does not contribute to its leading order as $a\to
0$, which is the same as in the two-dimensional case. Also
(\ref{transest2}) has to be replaced by the appropriate
two-dimensional estimate, which is again exponentially precise,
see \cite{EK3} for details.
\end{proof}

The technique used in these proofs can be applied to many other
cases. If $\Gamma$ is an infinite curve, the threshold of the
essential spectrum is moved and the estimates on
$\sharp\,\sigma_\mathrm{disc}(H_{\alpha,\Gamma})$ are no longer
relevant. On the other hand, the eigenvalue asymptotic
formula{\ae} remain valid under mild additional assumptions
\cite{EY2, EK3}:

\begin{theorem} \label{infin_asympt}
Suppose that $\gamma:\: \R\to\R^\nu,\: \nu=2,3\,$, satisfies
hypotheses of Theorems~\ref{dsexist} and \ref{ds3exist},
respectively. In addition, assume that $\dot k(s)$ and $\ddot
k(s)^{1/2}$ are $\OO(s^{-1-\epsilon})$ as $|s|\to\infty$, and
$\tau, \dot\tau \in L^\infty(\R)$ for $\nu=3$. Then the asymptotic
expansions from Theorems~\ref{loop2_asympt} and \ref{loop3_asympt}
hold for all the eigenvalues $\lambda_j(\alpha)$ of
$H_{\alpha,\Gamma}$, when $S_\Gamma := -{\D\over \D s^2} - {1\over
4}k(s)^2$ is now the operator on $L^2(\R)$ with the domain
$W^{2,2}(\R)$.
\end{theorem}
\begin{remark}
In this case we need not care about the multiplicity, because the
spectrum of $S_\Gamma$ in $L^2(\R)$ is simple. This may not be
true, of course, in the more general case to which the results
extend easily, namely for $\Gamma$'s consisting of disconnected
$C^4$ smooth edges, i.e. curves which do not touch or cross each
other. On the other hand, situation becomes considerably more
complicated in presence of angles or branchings; we will comment
on it in Sec.~\ref{ss: angle} below.
\end{remark}

\subsection{Surfaces in $\mathbb{R}^3$} \label{ss: 3strong}

The method works also for interactions supported by surfaces, but
the geometric part is naturally different. Let us consider first
the case of a $C^4$ smooth compact and closed Riemann surface
$\Gamma \subset \mathbb{R}^3$ of a finite genus $g$. In the usual
way \cite{Kli}, the geometry of $\Gamma$ is encoded in the metric
tensor $g_{\mu\nu}$ and Weingarten tensor $h_{\mu}\,^{\nu}$. The
eigenvalues $k_{\pm}$ of the latter are the principal curvatures
which determine the Gauss curvature $K$ and mean curvature $M$ by
 $$ 
K=\det (h_{\mu }\,^{\nu
})=k_{+}k_{-}\,,\quad M=\frac{1}{2}\,\mathrm{Tr\:} (h_{\mu
}\,^{\nu}) =\frac{1}{2}(k_{+}\!+k_{-})\,.
 $$ 
The operator $H_{\alpha,\Gamma}$ is defined as in Sec.~\ref{ss:
qf&bc}. For a compact $\Gamma$ the essential spectrum is
$[0,\infty)$ and we ask about the asymptotic behavior of the
negative eigenvalues as $\alpha\to \infty$. It will be expressed
again in terms of a comparison operator: the $S_\Gamma$ of
Theorem~\ref{loop2_asympt} has to be now replaced by
\begin{equation} \label{compar}
S_\Gamma:= -\Delta_{\Gamma} +K-M^2
\end{equation}
on $L^2(\Gamma,d\Gamma)$, where $\Delta_\Gamma = - g^{-1/2}
\partial_\mu g^{1/2} g^{\mu\nu} \partial_\nu$ is the
Laplace-Beltrami operator on $\Gamma$. The $j$-th eigenvalue
$\mu_j$ of $S_\Gamma$ is bounded from above by that of
$\Delta_\Gamma$ because
$$ K-M^2= -\frac{1}{4}(k_{+}-k_{-})^{2}\le 0 $$
in analogy with the curve case; in distinction to the latter the
two coincide when $\Gamma$ is a sphere. With these prerequisites
we can make the following claim:

\begin{theorem} \label{t:comp}
Under the stated assumptions, $\sharp\,\sigma_\mathrm{disc}
(H_{\alpha,\Gamma})\ge j$ for any fixed integer $j$ if $\alpha$ is
large enough. The $j$-th eigenvalue $\lambda_j(\alpha)$ of
$H_{\alpha,\Gamma}$ has the expansion
 $$ 
\lambda_j(\alpha) = -\frac{1}{4}\alpha^2 +\mu_j +\mathcal{O}(
\alpha^{-1} \ln\alpha)
 $$ 
as $\alpha\to\infty$, where $\mu_j$ is the $j$-th eigenvalue of
$S_\Gamma$. Moreover, the counting function $\alpha\mapsto
\sharp\,\sigma_\mathrm{disc} (H_{\alpha,\Gamma})$ behaves
asymptotically as
 $$ 
\sharp\,\sigma_\mathrm{d} (H_{\alpha,\Gamma}) =
\frac{|\Gamma|}{16\pi} \alpha^2 +\mathcal{O}( \alpha) \,,
 $$ 
where $|\Gamma|$ is the Riemann area of the surface $\Gamma$.
\end{theorem}
\begin{proof}[Sketch of the proof]
To employ the bracketing, we need to construct a family of layer
neighborhoods of $\Gamma$. Let $\{n(x):\, x\in\Gamma\}$ be a field
of unit vectors normal to the manifold; such a field exists
globally because $\Gamma$ is orientable. We define a map
$\mathcal{L}_a:\: \Gamma\times(-a,a) \to \mathbb{R}^3$ by
$\mathcal{L}_a(x,u)= x + un(x)$. Due to the assumed smoothness of
$\Gamma$ it is a diffeomorphism for all $a$ small enough, mapping
onto the sought layer neighborhood $\Omega_a= \{
x\in\mathbb{R}^3:\: \mathrm{dist}(x, \Gamma)<a\}$.

By bracketing we get a two-sided estimate for the negative
spectrum of $H_{\alpha,\Gamma}$ by means of the layer part
operators. The latter can be analyzed by means of the curvilinear
coordinates following \cite{DEK}, see \cite{Ex3} for details. One
arrives at estimates through operators with decoupled variables,
$S^{\pm}_a \otimes I + I\otimes T^{\pm}_{\alpha,a}$ with
$$ S_a^{\pm} :=
-C_{\pm}(a)\Delta_{\Gamma } +C_{\pm}^{-2}(a)(K-M^2)\pm va
$$
and the transverse part which is the same as in the proof of
Theorem~\ref{loop2_asympt}. Here $C_\pm(a):= (1\pm
a\varrho^{-1})^2$ with $\varrho:= \max (\{\left\|
k_{+}\right\|_\infty, \left\| k_{-}\right\|_\infty \})^{-1}$ and
$v$ is a suitable constant. The rest of the argument is again
analogous to Theorem~\ref{loop2_asympt}; to get the counting
function one has to employ the appropriate Weyl formula \cite{Ch}.
\end{proof}

 \begin{remark}
The connectedness assumption is made for simplicity; the claim
remains valid if $\Gamma$ is a finite disjoint union of $C^4$
smooth compact Riemann surfaces of finite genera. Moreover, the
asymptotic formula for $\sharp\,\sigma_\mathrm{d}
(H_{\alpha,\Gamma})$ is preserved if $\Gamma$ has a nonempty and
smooth boundary, see \cite{Ex3} for a more detailed discussion.
 \end{remark}

Under additional assumptions the technique can be applied also to
interactions supported by infinite surfaces. One possible set of
hypotheses looks as follows:
\begin{description}
\item[\emph{(as1)} injectivity] the map $\mathcal{L}_a:\:
\Gamma\times(-a,a) \to \Omega_a\subset\R^3$ defined above is
injective for all $a$ small enough.
\item[\emph{(as2)} uniform ellipticity] $\,c_{-}\delta_{\mu \nu}
\leq g_{\mu \nu}\leq c_{+}\delta_{\mu \nu}$ for some $c_{\pm}>0$.
\item[\emph{(as3)} asymptotic planarity] $\,K,M \rightarrow 0$ as
the geodesic radius $r\rightarrow \infty$.
\end{description}
One can also replace the last requirement by a stronger assumption
which implies, however, at the same time the validity of (as1):
\begin{description}
\item[\emph{(as3')} asymptotic direction] the normal vector
$\,n\rightarrow n_0\,$ as the geodesic radius $\,r\rightarrow
\infty\,$, where $n_0$ is a fixed vector.
\end{description}
Then one can prove in a similar way as above the following result
\cite{EK}:
\begin{theorem} \label{assmut}
(i) Assume (as1) and (as3), or alternatively (as3'), then we have
$\inf \sigma _{\mathrm{ess}}(H_{\alpha,\Gamma})=
\epsilon(\alpha)$, where $\epsilon(\alpha) +\frac14 \alpha^2=
\OO(\alpha^2 \e^{-\alpha a/2})$ as $\alpha\to \infty$. \\ [1mm]
(ii) In addition, assume (as2). Unless $\Gamma$ is a plane, there
is at least one isolated eigenvalue of $H_{\alpha,\Gamma}$ below
the threshold of the essential spectrum for all $\alpha$ large
enough, and moreover, the eigenvalues $\lambda _{j} (\alpha)$ of
$H_{\alpha,\Gamma}$ have the following asymptotic expansion,
 $$
 \lambda _{j}(\alpha)=-\frac{1}{4} \alpha ^{2} +\mu_{j}
 +\mathcal{O}(\alpha ^{-1}\ln\alpha )
 $$
as $\,\alpha\to\infty$, where $\mu_{j}$ the $j$-th eigenvalue of
the corresponding operator (\ref{compar}) counted with
multiplicity.
\end{theorem}

In addition to the eigenvalue expansion, we have established also
the existence of curvature-induced bound states for non-planar
$\Gamma$ with suitable spatial asymptotic properties in the
situation, when the particle is attracted to $\Gamma$ sufficiently
strongly.

\subsection{Periodic curves}

Other large class of leaky graph Hamiltonians for which we can
investigate the strong-coupling asymptotics in the described way
concerns periodic manifolds. Let us start with a planar curve
$\Gamma$ being the graph of a $C^4$ smooth function $\gamma:\:
\R\to\R^2$. In view of the smoothness, the signed curvature $k$ is
a $C^2$ function; we assume
\begin{description}
\item[\emph{(as1)} curvature periodicity] there is $L>0$ such that
$k(s+L)=k(s)$.
\item[\emph{(as2)} curve periodicity] $\int^{L}_{0} k(s)\,\D s=0$.
We may suppose that the normal at $s=0$ is $(1,0)$, then
$\Gamma(\cdot+L)-\Gamma(\cdot)=(l_{1},l_{2})$ where period-shift
components are $l_{j} := \int^{L}_{0}\sin \left( \frac{\pi}{2}
(2-j) -\int^{t}_{0}k(u)\,\D u \right)\,dt$. Again without loss of
generality, we may suppose that $l_1>0$.
\item[\emph{(as3)} period cell match] the map (\ref{strip map}) is
injective for all $a$ small enough and $\Phi_a((0,L)\times
(-a,a))\subset \Lambda:=(0,l_{1})\times {\mathbb R}$.
\end{description}
As usual in a periodic situation we have to perform the
Floquet-Bloch decomposition \cite{Ku3}. The operator
$H_{\alpha,\Gamma}(\theta)$ on $L^2(\Lambda)$ is for a
$\theta\in[-\pi,\pi)$ defined through the quadratic form as in
Sec.~\ref{ss: qf&bc}; its domain consists of functions $u\in
W^{1,2}(\Lambda)$ satisfying the boundary conditions $u(l_{1},
l_{2}+\cdot)=\e^{i\theta}u(0,\cdot)$. In a standard way \cite{EY2}
one proves existence of a unitary $\mathcal{U}: \,
L^2(\mathbb{R}^2) \to \int^\oplus_{[-\pi,\pi)} L^2(\Lambda)\,
\mathrm{d}\theta$ such that
$$ \mathcal{U} H_{\alpha,\Gamma} \mathcal{U}^{-1} =
\int^\oplus_{[-\pi,\pi)} H_{\alpha,\Gamma}(\theta)\,
\mathrm{d}\theta \quad\; \mathrm{and} \;\quad
\sigma(H_{\alpha,\Gamma}) = \bigcup_{[-\pi,\pi)}
\sigma(H_{\alpha,\Gamma}(\theta))\,; $$
since $\gamma((0,L))$ is compact we have
$\sigma_{\mathrm{ess}}(H_{\alpha,\Gamma}(\theta))=[0,\infty)$.
Next we need a comparison operator on the curve analogous to
$S_\Gamma$ of Theorem~\ref{loop2_asympt}. For a fixed $\theta\in
[-\pi,\pi)$ it is given by the same symbol,
 $$ S_{\Gamma}(\theta)=-\frac{\D^{2}}{\D s^{2}}-\frac{1}{4}k(s)^{2}
\quad\mathrm{on}\quad L^{2}((0,L))$$
with the domain $\{ u\in W^{2,2}((0,L)):\:
u(L)=\e^{i\theta}u(0),\: u^{\prime}(L)=
\e^{i\theta}u^{\prime}(0)\}$. We denote by $\mu_{j}(\theta)$ the
$j$-th eigenvalue of $S_{\Gamma}(\theta)$ counted with
multiplicity. Modifying the method of Sec.~\ref{ss: strong curve}
to the present situation we get following result \cite{EY2}:

 \begin{theorem} \label{strong_per_curve}
To any $j\in\mathbb{N}$ there is an $\alpha_j>0$ such that
$\sharp\,\sigma_\mathrm{disc}(H_{\alpha,\Gamma}(\theta))\geq j$
holds for $\alpha\ge \alpha_j$ and any $\theta\in [-\pi,\pi)$. The
$j$-th eigenvalue of $H_{\alpha,\Gamma}(\theta)$ counted with
multiplicity has the asymptotic expansion
 $$ \lambda_{j}(\alpha,\theta)=-\frac{1}{4}\alpha^{2}
+\mu_{j}(\theta)+\mathcal{O}(\alpha^{-1}\ln\alpha)
 $$
as $\alpha\to\infty$, where the error term is uniform with respect
to $\theta\in [-\pi,\pi)$.
 \end{theorem}

Combining this result with Borg's theorem on the inverse problem
for Hill's equation, we can make a claim about gaps of $\sigma
(H_{\alpha,\Gamma})$:
 \begin{corollary} \label{gap}
Assume that $\Gamma$ is not a straight line, $k\ne 0$, then the
spectrum of $H_{\alpha,\Gamma}$ contains open gaps for all
$\alpha$ large enough.
 \end{corollary}

In an exactly similar manner one can treat periodic curves in
$\R^3$ applying the technique to the fiber operator in the
Floquet-Bloch decomposition of $H_{\alpha,\Gamma}$ \cite{EK3};
more care is needed only when defining
$H_{\alpha,\Gamma}(\theta)$, since quadratic forms cannot be now
used.

\begin{theorem} \label{evfibe}
Let $\Gamma$ be a periodic curve, without self-intersections and
with the global Frenet frame, given by a $C^4$ smooth function
$\gamma:\:\R\to\R^3$. Suppose, in addition, that the period cells
$\Gamma_\mathrm{p}$ of $\Gamma$ and $\Lambda$ referring to the
corresponding operator $H_{\alpha,\Gamma}$ match in the sense that
$\Gamma_\mathrm{p} = \Gamma\cap\Lambda$. Then $\sigma
_\mathrm{disc} (H_{\alpha,\Gamma}(\theta))$ has the properties
analogous to those of the previous theorem, in particular, the
$j$-th eigenvalue of $H_{\alpha,\Gamma}(\theta)$ has the
asymptotic expansion of the form
$$
\lambda _{j}(\alpha ,\theta )=\zeta_{\alpha }+\mu_{j}(\theta )+
\mathcal{O}(\e^{\pi \alpha })\quad \mathrm{as} \quad \alpha
\rightarrow -\infty\,,
$$
where $\mu _{j}(\theta )$ is the $j$-th eigenvalue of
$S_{\Gamma}(\theta)$ and the error is uniform w.r.t. $\theta $.
\end{theorem}

\begin{remarks} \label{per_rem}
(i) Corollary~\ref{gap} has, of course, its three-dimensional
analogue. The number of gaps which can be open depends on the
shape of $\Gamma$. Notice that the operator
$\int^\oplus_{[-\pi,\pi)} S_{\Gamma}(\theta)\, \D\theta$ has
generically infinitely many open gaps, and in such a case the
corresponding $H_{\alpha,\Gamma}$ can have any finite number for
$\alpha$ large enough. \\
(ii) The assumption about a match between the periodic
decompositions of the curve $\Gamma$ and the corresponding
$H_{\alpha,\Gamma}$ may seem restrictive. One should realize,
however, the period cell need not be rectangular as one usually
supposes when periodic Schr\"odinger operators are considered.
What we actually need is a complete ``tiling'' of $\R^\nu$ by
domains with piecewise smooth boundaries. In the case $\nu=3$ such
``bricks'' need not even be simply connected: remember what your
grandmother was doing with her crotchet to get an example of a
curve which is topologically inequivalent to a line, or in other
words, you cannot disentangle it by any local deformation (you can
only unwind it by ``pulling the ends''). \\
(iii) The case we have discussed above, namely a single infinite
curve in $\R^\nu$ periodic in a given direction, is the simplest
possibility. In a similar way one can treat infinite families of
curves periodic in $r\le\nu$ directions, the only restriction is
that their components have to satisfy individually the listed
assumptions and the distances between them must have a uniform
positive lower bound. The case $r=\nu$ differs from $r<\nu$
because then the basic cell is precompact, and therefore the
spectrum of each $H_{\alpha,\Gamma}(\theta)$ is purely discrete.
\\
(iv) A particular situation occurs when a periodic $\Gamma$
consists of disjoint compact components. The asymptotic expansions
are valid again but now the eigenvalues of the fiber comparison
operator are independent of the parameter $\theta$ so we have
 $$ \lambda_{j}(\alpha,\theta)=-\frac{1}{4}\alpha^{2}
+\mu_{j}+\mathcal{O}(\alpha^{-1}\ln\alpha)
 $$
and the respective expansion in the three-dimensional case (when
the topology may be again nontrivial -- imagine a chain of
interlocked rings). Naturally, the chances to have open gaps in
this situation are generally better than in the connected case. \\
(v) The last comment concerns interpretation of these results.
Recall that the deviation of $\sigma(H_{\alpha,\Gamma})$, in the
negative part at least, from the one described by the comparison
operator is due to quantum tunneling. Hence it must be sensitive
to the appropriate parameter, i.e. the Planck's constant if we
reintroduce it into the picture. However, the operator $-h^2\Delta
-v\delta(x-\Gamma)$ is the $h^2$ multiple of (\ref{formal}) if we
denote $\alpha:= vh^{-2}$; in this sense therefore the obtained
asymptotic formul{\ae} represent a semiclassical approximation.
\end{remarks}

Let us mention one more consequence of these considerations
\cite{BDE}.

\begin{theorem} \label{strong_ac}
Suppose that the curve $\Gamma\subset\R^\nu$ satisfies the
assumptions of Theorems~\ref{strong_per_curve} and \ref{evfibe}
for $\nu=2,3$, respectively. If $\nu=2$ to any $\lambda>0$ there
is an $\alpha_\lambda>0$ such the spectrum of the operator
$H_{\alpha,\Gamma}$ is absolutely continuous in $(-\infty,
-\frac14\alpha^2+\lambda]$ as long as $\alpha>\alpha_\lambda$. The
same is true for $\nu=3$ with $-\frac14\alpha^2$ replaced by
$\zeta_\alpha$ provided $-\alpha>\alpha_\lambda$.
\end{theorem}
\begin{proof}[Sketch of the proof]
It is easy to check that $\,\{ H_{\alpha,\Gamma}(\theta):\:
\theta\in[-\pi,\pi) \}$ is a type A analytic family. The spectral
interval in question contains a finite number of eigenvalue
branches, each is a real analytic function which can be written
through one of the above asymptotic expansions. The functions
$\mu_j(\cdot)$ are nonconstant, hence the same is true for
$\lambda_j(\alpha,\cdot)$ provided $(-1)^\nu \alpha$ is large
enough.
\end{proof}

To appreciate this result recall that if the orbit space of the
operator $H_{\alpha,\Gamma}$ is compact --
cf.~Remark~\ref{per_rem}(iv) above -- there is a way to establish
the (global) absolute continuity of such operators \cite{BSS, SS},
while the situation with $r<\nu$ is more difficult; we will
mention related results in Sec.~\ref{ss: modul} below.

\subsection{Periodic surfaces}

The treatment of operators $H_{\alpha,\Gamma}$ corresponding to
periodic surfaces is similar and we describe it only briefly
referring to \cite{Ex3} for details. We consider discrete
translations of $\mathbb{R}^3$ generated by an $r$-tuple
$\{l_i\}$, where $r=1,2,3$. We decompose $\Gamma$, supposed to be
a $C^4$ smooth Riemann surface, not necessarily connected, and
$\R^3$ into period cells $\Gamma_\mathrm{p}$ and $\Lambda$
assuming again that they match mutually. The Floquet-Bloch
decomposition proceeds as above: we define the fiber operators
$H_{\alpha,\Gamma}(\theta)$ on $L^2(\Lambda)$ through quadratic
forms defined on functions satisfying the appropriate boundary
condition; after that we prove existence of a unitary
$\mathcal{U}: \, L^2(\mathbb{R}^3) \to \int^\oplus_{[-\pi,\pi)^r}
L^2(\Lambda)\, \mathrm{d}\theta$ such that
$$ \mathcal{U} H_{\alpha,\Gamma} \mathcal{U}^{-1} =
\int^\oplus_{[-\pi,\pi)^r} H_{\alpha,\Gamma}(\theta)\,
\mathrm{d}\theta \quad\; \mathrm{and} \;\quad
\sigma(H_{\alpha,\Gamma}) = \bigcup_{[-\pi,\pi)^r}
\sigma(H_{\alpha,\Gamma}(\theta))\,.
$$
The spectrum of $H_{\alpha,\Gamma}(\theta)$ is purely discrete if
$\,r=3$ while $\sigma_\mathrm{ess}(H_{\alpha,\Gamma}(\theta))=
[0,\infty)$ if $\,r=1,2$; the eigenvalues are continuous functions
of the quasi-momenta $\theta_\mu$.

As before we need a comparison operator. Its fibers act on
$L^2(\Gamma_\mathrm{p}, d\Gamma)$ being defined, for instance, by
the following prescription
 $$ 
S_\Gamma(\theta) := g^{-1/2} (-i\partial_\mu+\theta_\mu) g^{1/2}
g^{\mu\nu} (-i\partial_\nu+\theta_\nu) + K - M^2
 $$ 
with the domain consisting of $\phi \in
W^{1,2}(\Gamma_\mathrm{p})$ such that $\Delta_{\Gamma}\phi\in
L^2(\Gamma_\mathrm{p}, d\Gamma)$ satisfying periodic b.c. Since
$\Gamma_\mathrm{p}$ is precompact and the curvatures involved are
bounded, the spectrum of $S_\Gamma(\theta)$ is purely discrete for
each $\theta \in [-\pi,\pi)^r$; we denote the $j$-th eigenvalue,
counting multiplicity, as $\mu_j(\theta)$. In the same way as
above we get

\begin{theorem} \label{t:comp_per}
Under the stated assumptions the following claims are valid:
\\[1mm]
(a) Fix $\lambda$ as an arbitrary number if $r=3$ and a
non-positive one for $r=1,2$. To any $j\in\mathbb{N}$ there is
$\alpha_j>0$ such that $H_{\alpha,\Gamma}(\theta)$ has at least
$j$ eigenvalues below $\lambda$ for any $\alpha>\alpha_j$ and
$\theta \in [-\pi,\pi)^r$. The $j$-th eigenvalue
$\lambda_j(\alpha, \theta)$ has then the expansion
\begin{equation} \label{asympt_per}
\lambda_j(\alpha, \theta) = -\frac{1}{4}\alpha^2 +\mu_j(\theta)
+\mathcal{O}( \alpha^{-1} \ln\alpha)
\end{equation}
as $\alpha\to\infty$, where the error term is uniform with respect
to $\theta$. \\[1mm]
(b) If the set $\sigma(S):= \bigcup_{\theta \in
[-\pi,\pi)^r}\sigma(S_\Gamma(\theta))$ has a gap separating a pair
of bands, then the same is true for $\sigma (H_{\alpha,\Gamma})$
provided $\alpha$ is large enough.
\end{theorem}

\subsection{Magnetic loops} \label{ss: mgloop}

Up to now we considered systems without external fields, which
leaves out many situation of experimental interest. For instance,
one of the often studied features of mesoscopic systems are
persistent currents in rings threaded by a magnetic flux. For a
charged particle (an electron) confined to a loop $\Gamma$ the
effect is manifested by the dependence of the corresponding
eigenvalues $\lambda_n$ on the flux $\phi$ through the loop,
conventionally measured in the units of flux quanta, $2\pi\hbar
c|e|^{-1}$. The derivative $\partial \lambda_n/\partial \phi$
equals $-{1\over c}I_n$, where $I_n$ is the persistent current in
the $n$--th state. For the ideal loop, e.g., the eigenvalues in
absence of other than magnetic potential are proportional to
$(n+\phi)^2$ so the currents depend linearly on the applied field.
The question is what can we say when the confinement is of the
type discussed here realized through an attractive interaction on
the loop.

We add a homogeneous magnetic field with vector potential $A=
\frac{1}{2} B(-x_2,x_1)$ to our considerations and investigate the
Hamiltonian formally given by
 \begin{equation} \label{mg_Ham}
 H_{\alpha,\Gamma}(B):= (-i\nabla-A)^2 -\alpha
\delta(x-\Gamma)
 \end{equation}
in $L^2(\R^2)$. To define it properly we use quadratic form
analogous to (\ref{Hamform}),
 $$ 
 \EE_{-\alpha m,B}[\psi]=\left\| \left(-i\partial_x +{1\over 2}By
 \right)\psi \right\|^{2}
 +\left\|\left(-i\partial_y -{1\over 2}Bx\right)\psi\right\|^{2}
 -\alpha \int_{\R^2}|(I_m\psi)(x)|^{2}\,\D x
 $$ 
with the domain $W^{1,2}(\mathbb{R}^{2})$. It is straightforward
to check that the form is closed and below bounded; we identify
the self-adjoint operator associated to it with
$H_{\alpha,\Gamma}(B)$.

We use the same technique based on bracketing and estimating the
operator in the strip in suitable coordinates. We need again a
comparison operator, this time
$$ S_\Gamma(B) = -{\mathrm{d}\over \mathrm{d}s^2}
- {1\over 4}k(s)^2 $$
on $L^2(0,L)$ with $ \psi(L-)= \mathrm{e}^{iB|\Omega|}\psi(0+),\,
\psi'(L-)= \mathrm{e}^{iB|\Omega|}\psi'(0+)$, where $\Omega$ is
the area encircled by $\Gamma$. Using it we can state the
following result which establishes, in particular, the existence
of persistent currents on a leaky loop for $\alpha$ large enough.

\begin{theorem} \label{mgloop}
Let $\Gamma$ be a $C^4$-smooth curve without self-intersections.
For a fixed $j\in\mathbb{N}$ and a compact interval $I$ we have
$\sharp\,\sigma(H_{\alpha,\Gamma}(B))\ge j$ for $B\in I$ if
$\alpha$ is large enough, and the $j$-th eigenvalue behaves as
$$ \lambda_j(\alpha, B) = -\frac{1}{4}\alpha^2 +\mu_j(B)
+\mathcal{O}( \alpha^{-1}\ln\alpha)\,,
$$
where $\mu_j(B)$ is the $j$-th eigenvalue of $S_\Gamma(B)$ and the
error term is uniform in $B$. In particular, for a fixed $j$ and
$\alpha$ large enough the function $\lambda_j(\alpha, \cdot)$
cannot be constant.
\end{theorem}
\begin{proof}[Sketch of the proof]
The argument is closely similar to the analysis of fiber operators
in Theorem~\ref{strong_per_curve}, the magnetic flux replacing
Floquet parameter, with small technical differences for which we
refer to \cite{EY3}.
\end{proof}

\section{A discrete analogue}

It is often useful to investigate in parallel with leaky graphs
analogous discrete structures in which the attractive interaction
is supported by suitably arranged families of point interactions.
Let us briefly recall the basic notions, for more information and
a rich bibliography we refer to \cite{AGHH}. Consider a
set\footnote{For simplicity we use the symbol $Y$ both for the map
$I\to \R^\nu$ and its range.} of points $Y=\{y_n\}_{n\in
I}\subset\R^\nu,\: \nu=2,3$; if $I$ is infinite we suppose that
$Y$ can accumulate only at infinity. The operators of interest are
point-interaction Hamiltonians $H_{\alpha,Y}$, typically with the
same interaction ``strength'' at each point, which are defined by
means of the boundary conditions
\begin{equation} \label{pi_bc}
L_1(\psi,y_j) -\alpha L_0(\psi,y_j)=0\,, \quad j\in I\,,
\end{equation}
expressed in terms of the generalized boundary values
$$ L_0(\psi,y):= \lim_{|x-y|\to 0}\, {\psi(x)\over
\phi_d(x\!-\!y)}\,, \; L_1(\psi,y):= \lim_{|x-y|\to 0}
\bigl\lbrack \psi(x)- L_0(\psi,y)\, \phi_d(x-y) \bigr\rbrack\,, $$
where $\phi_d$ are the appropriate fundamental solutions, namely
$\phi_2(x)= -{1\over2\pi}\, \ln|x|$ and $\phi_3(x)=(4\pi|x|)^{-1}$
related to the free Green's functions (\ref{freeG}) and
(\ref{freeG3}), respectively. The resolvent of $H_{\alpha,Y}$ is
given by Krein's formula,
\begin{equation} \label{krein}
(-H_{\alpha,Y}-k^2)^{-1} = G_k + \sum_{j,j'\in I}
[\Gamma_{\alpha,Y}(k)]^{-1}_{jj'} \left(
\overline{G_k(\cdot\!-\!y_{j'})}, \cdot \right) G_k(\cdot\!-\!y_j)
\end{equation}
for $k^2\in\rho\left( H_{\alpha,Y}\right)$ with $\im k>0$, where
$\Gamma_{\alpha,Y}(k)$ is a closed operator (which is bounded in
our case) on $\ell^2(I)$ the matrix representation of which is
 $$ 
\Gamma_{\alpha,Y}(k):= \Big\lbrack (\alpha-\xi_d^k) \delta_{jj'} -
G_k(y_j-y_{j'} (1-\delta_{jj'}) \Big\rbrack_{j,j'\in I}\,,
 $$ 
where $\xi^k_d$ is the regularized Greens's function, $\xi^k_2=
-{1\over 2\pi}\, \left( \ln{k\over 2i} -\psi(1) \right)$ and
$\xi^k_3= {ik \over 4\pi}$. If $\alpha$ is independent of $j$, the
map $k\mapsto \Gamma_{\alpha,Y}(k)$ is analytic in the open upper
halfplane. Moreover, $\Gamma_{\alpha,Y}(k)$ is boundedly
invertible for $\im k>0$ large enough, while for $k\in\C^+$ not
too far from the real axis it may have a nontrivial null-space. By
(\ref{krein}) the latter determines the spectrum of the original
operator $H_{\alpha,Y}$ on the negative halfline in view the
following result analogous to Proposition~\ref{BS}.
\begin{proposition} \label{BS_pi}
(i) A point $-\kappa^2<0$ belongs to $\rho\left(H_{\alpha,Y}
\right)$ iff $\,\mathrm{ker\,} \Gamma_{\alpha,Y}= \{0\}$. \\
[1mm] (ii)
If the operator-valued function $\kappa\mapsto \Gamma_{\alpha,Y}
(i\kappa)^{-1}$ has bounded values in an open interval
$J\subset\R_+$ with the exception of a point $\kappa_0\in J$,
where $\dim\mathrm{ker\,} \Gamma_{\alpha,Y}(i\kappa)=n$, then
$-\kappa_0^2$ is an isolated eigenvalue of $H_{\alpha,Y}$ of
multiplicity $n$. \\ [1mm]
(iii) an eigenfunction of $H_{\alpha,Y}$ associated with such an
eigenvalue $-\kappa_0^2$ is equal to $\psi= \sum_{j\in I} d_j
G_{i\kappa_0}(\cdot-y_j)$, where $d=\{d_j\}$ solves the equation
$\Gamma_{\alpha,Y} (i\kappa_0)d=0$.
\end{proposition}

Let us now review discrete analogues of some results derived
above.

\subsection{Curved polymers} \label{ss: polymer}

The operator $H_{\alpha,Y}$ referring to a straight equidistant
array $Y$ is called a polymer model in \cite{AGHH}. Let us look
what happens if we abandon the straightness. We adopt the
following hypotheses:

\begin{description}
 \item[{\em (ad) analogue of (a1)--(a3) }] Let $Y=\{y_n\}_{n\in\Z}
 \subset\R^\nu,\: \nu=2,3$, be such that $|y_j-y_{j+1}| =\ell$ for
 some $\ell>0$. It implies $|y_j-y_{j'}| \le \ell\,|j-j'|$; we
 suppose that the inequality is sharp for some $j,j'\in\Z$. Next we
 assume that there is a $c_1\in(0,1)$ such that $|y_j-y_{j'}| \ge
 c_1\ell\,|j-j'|$, and moreover, that there are $c_2>0,\, \mu>\frac12$,
 and $\omega\in(0,1)$ such that the inequality
 $$ 
1-\, {|y_j-y_{j'}|\over|j-j'|} \le c_2 \left\lbrack
1+|j+j'|^{2\mu} \right\rbrack^{-1/2}
 $$ 
 holds if $(j,j')$ belongs to the sector $S_\omega$ of assumption
 (a2).
\end{description}

Recall first known facts \cite{AGHH} about a straight polymer,
$|y_j-y_{j'}| = \ell\,|j-j'|$ for all $j,j'\in\Z$. Its spectrum is
purely absolutely continuous and consists of two bands which may
overlap if $\alpha$ is not large enough negative. Its threshold
$E_\nu^{\alpha,\ell}$ is always negative; in the three-dimensional
case it is known explicitly,
 $$ 
E_3^{\alpha,\ell} = {1\over \ell^2} \left\lbrack
\ln\left(1+{1\over 2}\, \e^{-4\pi\alpha\ell} +
\e^{-2\pi\alpha\ell} \sqrt{1+ {1\over 4}\, \e^{-4\pi\alpha\ell}}
\right) \right\rbrack^2\,,
 $$ 
while for $\nu=2$ we have $E_2^{\alpha,\ell}=
-\kappa^2_{\alpha,\ell}$, where $\kappa_{\alpha,\ell}$ solves the
equation
 $$ 
\alpha + {1\over 2\pi} \left(\gamma- \ln2\right) = g_{i\kappa}(0)
 $$ 
with $g_k(\theta) := {1\over 2\pi} \lim_{N\to\infty} \left\lbrace
\sum_{n=-N}^N {1\over 2} \left\lbrack \left(n + {\theta\ell\over
2\pi} \right)^2\! - \left( k\ell\over 2\pi\right)^2
\right\rbrack^{-1/2}\!\! - \ln N \right\rbrace$.

\begin{theorem} \label{dsexist_pi}
Let $Y$ satisfy the assumptions (ad), then $\sigma_\mathrm{ess}
(H_{\alpha,Y})$ is the same as for the corresponding straight
polymer, and the operator $H_{\alpha,Y}$ has at least one isolated
eigenvalue below $E_\nu^{\alpha,\ell}$ for any $\alpha\in\R$.
\end{theorem}
\begin{proof}[Sketch of the proof]
The argument is a direct counterpart of that used in the proof of
Theorem~\ref{dsexist}. One checks that the perturbation is
sign-definite, pushing the spectrum of $\Gamma_{\alpha,Y}$ down,
and compact; the result then  follows by continuity and
Proposition~\ref{BS_pi}, cf.~\cite{Ex4} for more details.
\end{proof}

\subsection{Isoperimetric problem}

Let us mention also a discrete analogue of the problem discussed
in Sec.~\ref{ss: isoper}; we will think now of the point
interactions as of beads on a loop-shaped string. To be precise,
suppose the curve $\Gamma$ is the range of a function
$\,\gamma:\:[0,L]\to \mathbb{R}^\nu$ which is continuous,
piecewise $C^1$ and such that $\gamma(0)=\gamma(L)$, and
furthermore, $|\dot \gamma(s)|=1$ holds for any $s\in[0,L]$ for
which $\dot \gamma(s)$ exists. We consider the set $Y=\{y_j\}$
given by
 \begin{equation} \label{sites}
 y_j:= \gamma\left( \frac{jL}{N} \right)\,, \quad
 j=0,1,\dots,N-1\,,
 \end{equation}
with the indices regarded as integers, $y_j= y_{j
\mathrm{(mod\,}N\mathrm{)}}$. A distinguished element of the
described class is a regular polygon $\tilde{\mathcal{P}}_N$ for
which the points $y_j$ lie in a plane $\subset\mathbb{R}^\nu\,$
(this is trivial if $\nu=2$) at a circle of radius
$\frac{L}{N}\left( 2\sin \frac{\pi}{N}\right)^{-1}$.

We will suppose that the Hamiltonian $H_{\alpha,Y}$ has the
property analogous to (\ref{ex_ground}), namely that it possesses
a negative principal eigenvalue $\lambda_1(\alpha,Y)$; this is
automatically satisfied if $\nu=2$ while for $\nu=3$ it is true if
$-\alpha$ is large enough \cite{AGHH}. A counterpart to
Theorem~\ref{isoper} now reads
 \begin{theorem} \label{pi_isoper}
Under the stated assumptions the eigenvalue
$\lambda_1(\alpha,Y_\Gamma)$ is for a fixed $\alpha$ and $L>0$
globally sharply maximized by a regular polygon, $\Gamma=
\tilde{\mathcal{P}}_N$.
 \end{theorem}

We need a geometric result analogous to Proposition~\ref{chord}
which was proved in different ways in \cite{Lu, Ex6}, see also
\cite{Ex5} for a local proof.
 \begin{proposition} \label{disc_chord}
Let the set $Y$ be such that $|y_{j+1}-y_j|\le\frac{L}{N}$, then
for any $k$ and $p\in(0,2]$ the following inequality is valid
  $$
 \sum_{j=1}^N |y_{j+k}-y_j|^p\,
 \le\, \frac{N^{1-p}L^p \sin^p \frac{\pi k}{N}}{\sin^p
 \frac{\pi}{N}}\,
  $$
 \end{proposition}
Notice that the right-hand side is now the value of the sum for
the regular polygon, and that a similar reverse inequality holds
for negative powers $p\in[-2,0)$.

\begin{proof}[Sketched proof of Theorem~\ref{pi_isoper}]
The argument is analogous to that in the proof of
Theorem~\ref{isoper}. All the elements are in place: the ground
state is non-degenerate and for the regular polygon it has a
symmetry, this time with respect to a discrete group of rotations
which implies that the corresponding eigenfunction of
$\Gamma_{\alpha,Y}(i\kappa)$ is $N^{-1/2}(1,\dots,1)$. Using it to
make a variational estimate and employing the strict convexity and
monotony of the resolvent kernel we find that the chord-sum
inequality of Proposition~\ref{disc_chord} has to be valid with
$p=1$; a detailed account of the proof can be found in \cite{Ex5}.
\end{proof}

\subsection{Approximation by point interactions} \label{ss: approx}

Discrete ``leaky graphs'' described here are useful not only as
mathematical objects analogous to the main topic of this review.
As we are going to mention now, suitable families of them can be
used to approximate the ``true'' leaky graphs. Importance of such
an approximation stems from the fact that apart of particular
cases where symmetry allows for separation of variables we have no
efficient method to find spectral properties of the operator
$H_{\alpha, \Gamma}$. It is true that Proposition~\ref{BS} makes
it possible to rephrase the original PDE problem as solution of an
integral equation of Birman-Schwinger type but this is in general
a task which not easy either; the purpose of the approximation is
to convert it into an essentially algebraic problem.

To get an idea how to proceed in constructing the approximation
one can compare the spectra of $H_{\alpha,\Gamma}$ corresponding
to a straight line $\Gamma$ to that of a straight polymer
mentioned in Sec.~\ref{ss: polymer} above. We let the spacing
between point interactions go to zero. If the two spectra should
coincide in the limit the coupling parameter must be inversely
proportional to the spacing. It looks queer at a glance but one
has to keep in mind that the coupling described by the boundary
conditions (\ref{pi_bc}) becomes \emph{weaker} as $\alpha$
increases. We have the following result:
\begin{theorem} \label{th: aproxx_pi}
Let $\Gamma\subset\R^2$ be a finite graph obeying the assumptions
(g1), (g2) and $\alpha>0$. Choose $k$ with $\im k>0$ such that the
equation $\sigma -\alpha R^k_{m,m}\sigma = \alpha R^k_{m,\D
x}\psi$ has for any $\psi\in L^2(\R^2)$ a unique solution $\sigma$
which has a bounded and continuous representative on $\Gamma$.
Suppose next that there is a family $\{Y_n\}_{n=1}^\infty$ of
non-empty finite subsets of $\Gamma$ such that $|Y_n|:=\sharp\,Y_n
\to \infty$ and the following relations hold
  $$ 
{1 \over |Y_n|} \sum_{y \in Y_n} f(y) \: \to \: \int_{R^2} f(x)\,
\D m = \int_\Gamma f(s)\, \D s
  $$ 
for any bounded continuous function $f:\:\Gamma\to\C$, and
furthermore
  \begin{eqnarray*} 
&& \sup_{n \in \N} {1 \over |Y_n|} \sup_{x \in Y_n} \sum_{y \in
Y_n \setminus \{x\} } G_k (x-y) < \frac{1}{\alpha|\Gamma|}\,,
\\ \label{hypothesis 3}
&& \sup_{x \in Y_n} \bigg| {1 \over |Y_n|} \sum_{y \in Y_n
\setminus \{x\} } \sigma(y) G_k (x-y) - (R^k_{\D x,m} \sigma)(x)
\,\bigg| \: \to \: 0
  \end{eqnarray*}
as $n \to \infty$, where $|\Gamma|$ is the sum of all the edge
lengths in $\Gamma$. Then the family of the operators
$H_{\alpha_n,Y_n}$ with $\alpha_n:= |Y_n| (\alpha|\Gamma|)^{-1}$
approximates the leaky-graph Hamiltonian, $H_{\alpha_n,Y_n} \to
H_{\alpha,\Gamma}$ in the strong resolvent sense as $n\to\infty$ .
\end{theorem}
\begin{proof}[Sketch of the proof]
The argument is straightforward even if executing it needs some
effort. We have on one hand the resolvent of $H_{\alpha, \Gamma}$
given by Proposition~\ref{BS}, on the other hand the resolvent of
the approximating point-interaction Hamiltonians given by
(\ref{krein}); one has to show that their difference applied to
any $\psi\in L^2(\R^2)$ tends to zero as $n\to\infty$, see
\cite{EN1} for details.
\end{proof}

\begin{remarks} \label{pi_aprox_rem}
(i) A similar result holds for approximations of leaky surfaces in
$\R^3$ by families of three-dimensional point interaction, cf. the
paper \cite{BFT} where such an approximation was studied for the
first time. Both the two-dimensional and three-dimensional results
are valid for more general sets $\Gamma$ and couplings
which are non-constant functions over such a $\Gamma$. \\
(ii) Validity of the theorem extends to two-dimensional systems
exposed to a magnetic field perpendicular to the plane. The field
need not be homogeneous but it has to be sufficiently smooth \cite{Oz}. \\
(iii) As a mean for numerical calculations the above theorem is
not optimal. The convergence of eigenvalues is slow, roughly as
$\OO(n^{-1/2})$, and only in the strong resolvent sense. These
flaws can be overcome by considering the family
$\epsilon^2\Delta^2 +H_{\alpha_n,Y_n}$ which converges to
$\epsilon^2\Delta^2 + H_{\alpha,\Gamma}$ in the norm-resolvent
sense, and the limit approximates $H_{\alpha, \Gamma}$ in the same
sense as $\epsilon\to 0\:$ \cite{BO}. In addition, the numerical
approximations obtained by this method converge faster.
\end{remarks}

\subsection{Edge currents in the absence of edges} \label{ss:
edge}

Magnetic system exhibit interesting transport properties
manifested, in particular, through edge currents discovered in
\cite{Ha, MDS} and studied in numerous subsequent papers. The
effect is very robust and can be observed also in situations when
the ``edge'' is thin indeed consisting of just an array of point
interactions. Consider the situation when the interaction sites
are arranged equidistantly along a line, for which we can without
loss of generality take the $x$--axis in $\R^2\ni x=(x_1,x_2)$,
and a charged particle interacting with them is exposed to a
homogeneous magnetic field $B$ perpendicular to the plane. In such
a situation it is natural to use the Landau gauge in which the
Hamiltonian can be formally written as
 $$ 
(-i\partial_{x_1} +Bx_2)^2 -\partial_{x_2}^2 +\sum_j \tilde\alpha
\delta(x_1\!-\!x_1^{(0)}\!-\!j\ell)\,,
 $$ 
where $\ell$ is the interaction sites spacing; we write
$\tilde\alpha_j$ to stress that this formal constant is not
identical with the ``true'' coupling parameter which enters the
boundary conditions analogous to (\ref{pi_bc}) in a proper
definition of the operator.

The first thing to observe is that the infinitely degenerate
eigenvalues which the Hamiltonian has in the absence of the point
interactions are preserved in the spectrum, because one can
construct eigenfunctions vanishing at $(\!x_1^{(0)}\!
+\!j\ell,0),\: j\in\Z$. Since the system is $\ell$ periodic, one
can perform Floquet decomposition and analyze the fiber operators
on the strip, which plays here the role of period cell, with a
single point interaction by Krein's formula analogous to
(\ref{krein}), arriving thus at the following conclusion
\cite{EJK}:

\begin{theorem} \label{noedge}
For any $\alpha\in\R$ the spectrum of the indicated operator
consists of the Landau levels $B(2n\!+\!1), \: n=0,1,2,\dots$, and
absolutely continuous spectral bands; between each adjacent Landau
levels there is one such band and the lowest one lies below $B$,
the unperturbed spectral threshold.
\end{theorem}

Moreover, one can compute the probability current associated with
the generalized eigenfunction for a fixed value of the
quasi-momentum to see that it is a nontrivial vector field
describing transport along the array \cite{EJK}. If the point
interaction are arranged along a non-straight line, an explicit
solution is no longer possible but the effect persists -- see,
e.g., \cite{ChE2} for a regular polygon arrangement. This is
interesting in connection with the result mentioned in
Remark~\ref{pi_aprox_rem}(ii): arranging point interactions
densely around the loop we can approximate the operators discussed
in Sec.~\ref{ss: mgloop}, the ``edge'' currents are then nothing
but the persistent currents considered there in the particular
case of a strong coupling.

\section{Other results}

\subsection{Periodically modulated wires} \label{ss: modul}

In Remark~\ref{rem1def}(i) we have mentioned that the definition
extends to the situation with a non-constant coupling referring to
the formal expression (\ref{formal}). Similarly we proceed in the
case of codimension two; one has to replace $\alpha$ in the
boundary condition of (\ref{cod2bc}) by $\alpha(s)$. We will
denote such operators again $H_{\alpha,\Gamma}$ where we have now
$\alpha:\:\R\to(0,\infty)$ for $\nu=2$ and $\alpha:\:\R\to\R$ for
$\nu=3$. The following result is of a particular
interest\footnote{Recall that for a non-straight periodic $\Gamma$
and a constant $\alpha$ we proved the absolute continuity only at
the bottom of the spectrum provided the interaction is strong
enough, cf.~Theorem~\ref{strong_ac}.}:

\begin{theorem} \label{ac_straight}
Suppose that $\alpha\in L^\infty(\R)$ is a periodic function, then
the spectrum of $H_{\alpha,\Gamma}$ is purely absolutely
continuous; its negative part is non-empty and consists of at most
finite number of bands.
\end{theorem}
\begin{proof}[Sketch of the proof]
The argument is based of investigation of the scattering for the
pair $(H_{\alpha,\Gamma},-\Delta)$. Using the Floquet
decomposition one establishes the existence of wave operators for
the fibers and the limiting absorption principle; the method
employs complexification of the quasi-momentum \`{a} la Thomas. To
fill the details in the case $\nu=2$ one has to proceed as in
\cite{Fr1, Fr2, FS}\footnote{For a recent more general result in
arbitrary dimension see \cite{FS2}.} -- note that the analogous
result for a regular potential ``ditch'' was derived in \cite{FKl}
-- the full proof for $\nu=3$ can be found in \cite{EF}.
\end{proof}

 Let us mention also that in the case $\nu=3$ one has an analogous
 result in the situation when a periodic interaction is supported
 by an infinite family of parallel lines arranged equidistantly in
 a plane $\subset\R^3\,$, cf.~\cite{EF}.

\subsection{A line--and--points model} \label{ss: line_point}

The number of explicitly solvable models in this area is not
large, in particular, if we exclude those which can be treated by
separation of variables. For instance, Theorems~\ref{dsexist} and
\ref{geneigenvTH} tell us about the discrete spectrum and
scattering due to a local perturbation of a straight leaky wire
but it is difficult to find the eigenvalues or the on--shell
S-matrix for a particular shape of the deformation. One can
achieve more in a caricature model in which a straight line is
perturbed just by a finite family of point interactions, so the
Hamiltonian in $L^2(\R^2)$ can be formally written as
 $$ 
 -\Delta -\alpha \delta (x-\Sigma) +\sum_{i=1}^n \tilde\beta_i
 \delta(x-y^{(i)})\,,
 $$ 
where $\Sigma :=\{(x_1,0);\:x_1\in\R\},\; \alpha>0$, and
$Y:=\{y^{(i)}\}$ is the perturbation support; we use again tilde
to stress that the $\tilde\beta_i$ are not the ``true'' coupling
parameters appearing in the boundary conditions analogous to
(\ref{pi_bc}).

To analyze such an operator, denoted $H_{\alpha,\beta}$, properly
defined, we need its resolvent. It can be expressed in a way
similar to Proposition~\ref{BS} with the auxiliary Hilbert space
being now $L^2(\R)\oplus\C^n$, or alternatively in analogy with
Theorem~\ref {res_scatt}, i.e. by Krein's formula as a rank $n$
perturbation to the resolvent of $H_{\alpha,\Sigma}$; we refrain
from stating the explicit formul{\ae} which can be found with
proofs in \cite{EK6}. Due to the finite rank of the perturbation
the analysis reduces to an essentially algebraic problem. This
allows us to prove various results, in particular

\begin{theorem} \label{line_point_spectrum}
Let $\beta=(\beta_{1},...,\beta_{n})\in\R^n$ and $\alpha >0$. The
operator $H_{\alpha,\beta}$ has a non-empty discrete spectrum,
$1\le \sharp\, \sigma(H_{\alpha,\beta})\le n$; the number of
eigenvalues is exactly $n$ if all the numbers $-\beta_{i}$ are
large enough. In particular, if $n=1$ and $y=(0,a)$, there is a
single eigenvalue; it tends to $-\frac14\alpha^2$ as
$|a|\to\infty$ when $\zeta_{\beta}:=-4 \e^{2(-2 \pi \beta +\psi
(1))} \in (-\frac14\alpha^2,0)$ and to $\zeta_\beta$ in the
opposite case.
\end{theorem}
Recall the usual definition of a resonance as a pole of the
analytical continuation of the resolvent across (a part of) the
continuous spectrum to the lower half-plane.
\begin{theorem} \label{line_point_res}
Let again $n=1$ and $y=(0,a)$, and suppose that $\zeta_{\beta}>
-\frac{1}{4}\alpha^2$. Then for all $|a|$ large enough the
Hamiltonian has a unique resonance, $z(a)=\mu(a)-i\nu(a)$ with
$\nu(a)>0$, which in the limit $|a|\to\infty$ behaves as
 $$ 
 \mu(a)=\zeta_{\beta}+\mathcal{O}(\e^{-a \zeta_{\beta}})\,, \quad \nu(b)
 =\mathcal{O}(\e^{-a \zeta_{\beta}})\,.
 $$ 
\end{theorem}

We will not give details of the proofs referring to \cite{EK6} for
a full exposition. In this paper also other spectral and
scattering properties of this system are discussed as well as an
extension to the three-dimensional situation with the interaction
supported by a plane and a finite number of points. Furthermore,
resonances in the case $n>1$ are discussed in \cite{EIK} where one
analyzes also the decay of an eigenstate of $H_{0,\beta}$ due to
the presence of the leaky line -- see also Sec.~\ref{ss: time}.

\subsection{Numerical results} \label{ss: num}

The emphasis in this review is on analytic approaches to the
problem, hence numerical methods will be mentioned only briefly.
As we have mentioned the cases in which the spectral problem can
be solved by  separation of variables are rare and in a sense
trivial. Sometimes, however, one can divide $\R^\nu$ into regions
exterior to $\Gamma$ in which such a separation is possible and to
find the solutions by the method known in physics as ``mode
matching''. Consider, e.g., a circular $\Gamma$ with an
angle-dependent coupling; one can use inside and outside the
circle the Ans\"atze $\sum_{m\in\Z} c_m^{(\pm)} f_m^{(\pm)}(r)\,
\e^{im\varphi}$ where $f_m^{(\pm)}$ are suitable Bessel functions
and to find the coefficients $c_m^{(\pm)}$ using the boundary
conditions which couple the two parts of the plane -- see
\cite{ETa} for this and similar examples.

The most versatile method, however, is based on the approximation
of leaky-graph Hamiltonians by point interactions as we have
discussed in Sec.~\ref{ss: approx}; examples of spectra obtained
in this way can be found in \cite{EN2, EN1, Oz}. A case of
particular interest concerns resonances for $H_{\alpha,\Gamma}$
when $\Gamma$ is a straight line with a buckling having a narrow
bottleneck. Of course, the approximation method applies to finite
$\Gamma$'s only but one can use the approach popular among the
physicists --- rigorously justified so far only for
one-dimensional potential scattering \cite{HM} --- in which one
cuts the system to a finite length and observes the dependence of
eigenvalues on the cut-off size; resonances are manifested by a
pattern in which intervals of almost constancy are interlaced with
steep jumps. In our case resonances appear when the bottleneck
half-width becomes comparable with the characteristic transverse
size of the generalized eigenfunction (\ref{gensigma}), i.e.
$(2\alpha)^{-1}$, cf.~\cite{EN1}.

\section{Open problems}

While we have reviewed numerous results in the previous sections,
the topic we are discussing here is new and many questions remain
still open. In this closing section we are going to list some
problems the reader may wish to learn about, or better, to attack.
As usual in mathematics, the number of generalizations is
unlimited and we restrict only to those which we regard as both
meaningful and reasonably close to the above exposition. Even in
such a class one can find very different problems. Some of those
listed below are rather technical and require mostly perspiration
to achieve the result; others are more challenging when we do not
know what the result might be, which method to apply, or possibly
neither of these -- in short, one cannot start without a proper
inspiration. Presenting the problems, however, we will not rank
them according to these criteria but rather list them in the order
they appear in the text.

\subsection{Approximation by regular potential ``ditches''}

This is an important link between our singular model and
description of realistic systems of ``wires'' of a small but
finite width\footnote{In the photonic-crystal setting, this
problem is discussed in \cite{FK2}. The relation between the leaky-graph Hamiltonian and the corresponding pseudo-differential (Dirichlet-to-Neumann) operator is another foundational issue which deserves attention, especially in view of the motivating numerical results obtained for particular geometries \cite{KK2}.}. In
Theorem~\ref{ditch_approx} we considered the simplest case of a
single $C^2$ curve $\Gamma\subset\R^2$. If $\Gamma$ has angles, or
even it is a nontrivial graph with branchings, one expects the
analogous result to be valid, however, the method used in the
proof is no longer applicable and it seems likely that the edge
and vertex parts of the approximating potentials have to be
treated separately. Similar results can be expected if the
``edges'' of $\Gamma$ are surfaces in $\R^3$.

A related question concerns existence of curvature-induced bound
states in such ``fat'' leaky graphs, as well as counterparts of
other results discussed in Sec.~\ref{s: geom}. Using the above
mentioned convergence result, combined possibly with minimax
estimates, one can establish the analogue of Theorem~\ref{dsexist}
provided the potential ``ditches'' in question are sufficiently
deep and narrow; it is less clear whether the claim will remain
true generally outside the asymptotic regime.

On the other hand, a potential approximation to
$H_{\alpha,\Gamma}$ is more delicate in case of codimension two. A
hint can be obtained from approximations of a two-dimensional
point interaction \cite{AGHH} where one starts from a potential
well having a zero-energy resonance and scales it in a particular
nonlinear way. One can conjecture that this would yield an
approximation for a single smooth $\Gamma$ if we take such a
potential family in the normal plane to $\Gamma$, while the
problem may be more complicated in presence of angles and
branchings.

\subsection{More singular leaky graphs}

The previous problem brings to mind a related question. Transverse
to the graph edges we choose the $\delta$ interaction as the mean
to describe the way in which the particle is attracted to
$\Gamma$. This is not the only possibility, though, even if we
restrict ourselves to interactions supported by a single point. We
know that such singular interactions form a four-parameter family
containing some prominent cases such as the $\delta'$ interaction
\cite{AGHH}. Let us recall that the latter are not mere
mathematical artefacts as it can be seen, e.g., from the fact that
they are approximated by triples of properly scaled $\delta$
interactions \cite{CS} and even by regular potentials \cite{AN,
ENZ}.

It is appropriate at this place to mention that we have not
discussed here counterparts of leaky-graph Hamiltonians with a
\emph{repulsive} interaction, formally $-\Delta
+\alpha\delta(x-\Gamma)$ with $\alpha>0$, which can be introduced
in the case $\mathrm{codim\,}\Gamma=1$ as in Sec.~\ref{ss: qf&bc};
the reason is that the model has different spectral and scattering
properties as well as physical interpretation. A combination of
the attractive and repulsive interactions on a ``triplicated''
graph where each edge of $\Gamma$ is replaced by three close edges
approximating in the cut the $\delta'$ interaction according to
\cite{CS} makes perfect sense, however, and justifies interest in
such models.

For $\delta'$ and more general leaky graphs we do not have at our
disposal a ``natural'' quadratic form definition analogous to
(\ref{Hamform}) or a Birman--Schwinger--type expression for the
resolvent, hence it is a priori not clear which ones of the
spectral and scattering properties discussed here can be extended
to this case.

\subsection{More on curvature induced spectra}

The existence of a discrete spectrum for an infinite curve
$\Gamma$ which is not straight but it is asymptotically straight
in a suitable sense as in Theorem~\ref{dsexist} is an interesting
result, however, a full understanding requires more. We already
know that $\sharp\,\sigma_\mathrm{disc}(H_{\alpha,\Gamma})$ can be
made arbitrarily large finite for a curve as simple as
$H_2(\beta)$ of Sec.~\ref{ss: star}. It is intuitively clear that
a rich discrete spectrum is to be expected when the edges of
$\Gamma$ come close to each other at numerous places or over long
stretches; it would be desirable to have a more quantitative
expression of this intuitive statement.

It is useful to stress at this place that the leaky curves have a
lot in common with \emph{quantum waveguides} \cite{ES, DE}. In
both cases we study solutions to the Schr\"odinger equation
localized in a tubelike region, here in a ``soft'' way through an
attraction to a curve in contrast to the ``hard'' way with
Dirichlet boundary conditions. This analogy makes is easier to
understand effects like existence of localized solutions due to
curvature because they are similar in both cases\footnote{This is
not to say that all properties are the same. As example is
provided by nodal lines of the discrete--spectrum eigenfunctions:
in thin waveguides they can be described by means of the
one-dimensional comparison operator \cite{FKr} while for leaky
curves they extend over the plane and it is more complicated to
analyze their behavior.}. This analogy suggests various questions,
for instance, about small bending asymptotic behavior. Take a
curve which is straight outside a compact and differs only
slightly from the straight line; an archetype of such a behavior
is a broken line corresponding to $H_2(\pi-\theta)$. Using the
method from the proof of Theorem~\ref{dsexist} one can check that
for $\theta$ small enough such a $H_{\alpha,\Gamma}$ has a single
eigenvalue $\lambda(\theta)$. We conjecture that
 \begin{equation} \label{phi^4}
 \lambda(\theta) = -\frac14 \alpha^2 - c\theta^4 + \OO(\theta^5)
 \end{equation}
in analogy with the slightly curved waveguide \cite{ABGM, DE}. The
positive constant $c$ here depends on $\alpha$, in particular, for
$H_2(\pi-\theta)$ we have $c=c' \alpha^2$. The methods discussed
in this paper, however, do not give a way to prove the conjectured
asymptotic behavior and another approach is needed.

The assumption (a2) in Sec.~\ref{ss: nonstr} was used to ensure
that the curvature induced spectrum is discrete. The proof of
Theorem~\ref{dsexist} shows, however, that any departure from a
straight line pushes the spectrum down\footnote{Or alternatively,
for a pure mathematician, to the left --- the same for the
previous text.}, the question is which character does it have
below $-\frac14 \alpha^2$. The cases to be analyzed include, e.g.,
curves with a slow curvature decay or a line with sparse
deformations. Analogies with one-dimensional Schr\"odinger
operators suggest that one might get different spectral types,
however, they have to be used with caution since the other
dimension may play a role\footnote{Interesting effects can be seen
also in the positive part of the spectrum. Recall the situation
where $\Gamma$ is an infinite family of concentric, equidistantly
spaced circles, then the $\sigma_\mathrm{ess}(H_{\alpha,\Gamma})$
consists of interlaced intervals of absolutely continuous and
dense pure point spectrum \cite{EFr}.}.

The above question belongs to those which are open also in the
quantum waveguide setting mentioned above. Another problem of this
sort concerns the spectral multiplicity, specifically, one would
like to know whether -- or possibly under which conditions -- is
the discrete spectrum of Theorem~\ref{dsexist} simple.

\subsection{More on star graphs} \label{pb: graph}

The example discussed in Sec.~\ref{ss: star} is simple,
nevertheless, some open questions remain. Notice that the
important parameters are the angles only, because a star-shaped
$\Gamma$ is self-similar and a change of $\alpha$ is equivalent to
a modification of the length scale. One can ask, for instance,
about the configuration which maximized the ground state; a
natural conjecture is that this happens in the case of the maximum
symmetry, $\beta_i=2\pi/N,\, i=1,\dots,N$. Less obvious is the
answer to the question about existence of closed nodal lines which
are expected if $N$ is large enough. A numerical example
\cite{EN1} based on Theorem~\ref{th: aproxx_pi} suggests this
happens, e.g., with the fourth eigenfunction of $H_{10}(\pi/5)$.
One can ask what is the minimum $N$ for which a leaky-star
Hamiltonian has a eigenfunction with a closed nodal line, about
stability of nodal patterns w.r.t. the angles, etc.

\subsection{Bound states for curved surfaces}

The analogy with quantum wave\-guides extends to the case of a
surface in $\R^3$: by Theorem~\ref{assmut} we know that non-planar
strongly attractive surfaces satisfying (as1)--(as3), or (as2) and
(as3') give rise to a non-empty discrete spectrum. Recall that for
Dirichlet quantum layers we also do not have an universal result;
they bind if their width is small enough, if the total Gauss
curvature is non-positive, or under various symmetry assumptions
\cite{DEK, CEK}. The thinness assumption is the analogue of a
strong coupling, $\alpha\to\infty$, discussed here. One is
interested whether and under which conditions does the curvature
imply existence of bound states for a fixed coupling parameter
$\alpha$.

Guided again by the quantum layer analogy \cite{EKr} one can ask
about the weak-coupling asymptotics of bound states corresponding
to mildly curved surfaces. In distinction to the conjecture
(\ref{phi^4}) one guesses that in the two-dimensional case the
binding will be exponentially weak with respect to a suitable
deformation parameter. The method to prove this result is not
clear, though.

\subsection{Bound states of nontrivial graphs of codimension two}

We have seen repeatedly that leaky graphs of codimension two
require more subtle analysis. This applies already to the
definition of $H_{\alpha,\Gamma}$ which we have briefly described
in Sec.~\ref{ss: cod2} for the case when $\Gamma$ is a curve.
There is little doubt that the boundary conditions (\ref{cod2bc})
can be used for a non-trivial graphs $\Gamma$ as well, however,
the proof using the resolvent formula (\ref{3BS}) has to be worked
out properly.

Another question concerns the possible analogue of
Corollary~\ref{dsgraph}. Due to the absence of a ``natural''
quadratic form associated with $H_{\alpha,\Gamma}$ it is not
obvious that bound states are preserved when edges are added to a
non-straight curve creating a non-trivial graph, although one
expects that this will be the case.

\subsection{Geometric perturbations for general graphs}

The results discussed in Sec.~\ref{ss: geom_pert} are formulated
for a single smooth curve or manifold. With an additional effort
one can extend validity of the asymptotic formul{\ae} to a general
graph $\Gamma$ provided the point at which the hiatus is centered
is in the interior of an edge. While a similar claim may be still
true for edges ``narrowly disconnected'' in a vertex, a more
subtle analysis is needed to see whether such a claim is valid.

\subsection{More on isoperimetric problems}

Theorem~\ref{isoper} raises naturally various questions about
possible extensions. A modification for a loop in $\R^3$ is
relatively straightforward and one can ask also whether there are
local maxima for loops of a fixed knot topology. On the other
hand, the corresponding problem for a closed surface in $\R^3$
seems to be more complicated and direct generalizations do not
work. Returning to the two-dimensional situation, and referring to
Sec.~\ref{pb: graph} above, one can ask more generally about
configurations which maximize the ground state for $\Gamma$ with a
fixed topology and edge lengths, etc.

\subsection{More on absolute continuity}

The case of $H_{\alpha,\Gamma}$ with a periodic curve $\Gamma$
remains to be one of the important challenges in this area. One
certainly expects that the spectrum of such an operator will be
purely absolutely continuous, however, proof of this assertion is
missing. So far we have only the result in the related but
different case expressed by Theorem~\ref{ac_straight}, and a
partial answer to the question given by Theorem~\ref{strong_ac};
let us note that the absolute continuity is expected be valid even
for $\Gamma$ consisting of infinitely many disconnected finite
components for which the strong-coupling argument does not work.

\subsection{More on scattering}

The setting we choose in Sec.~\ref{ss: scatt} has the advantage of
simplicity offered by the straight-line comparison operator
$H_{\alpha,\Sigma}$. Scattering for a general leaky graph with
straight leads outside a compact can be treated similarly, but
needs a more elaborate formulation: one has to compare motion on
each external lead with that of $H_{\alpha,\Sigma}$ using a
suitable identification operator. The situation becomes even more
complicated when the leads are not straight but only
asymptotically straight, in particular, if the ``curvature decay''
is slow so the curvature-induced interaction is of a long range.

A different class of problems concerns scattering at positive
energies. Here one has to distinguish two cases. If $\Gamma$ is
compact we deal with standard scattering on a potential, albeit a
singular one; this situation was discussed in \cite{BT}. On the
other hand, if there are infinite edges the situation is more
complicated due to presence of guided states; note that they can
exist also at positive energies, just take (\ref{gensigma}) with
$\lambda>0$. This case remains so far largely untreated.

\subsection{Strong coupling behavior of scattering}

Speaking about scattering one may ask whether there are asymptotic
formul{\ae} analogous to those we derived for the discrete
spectrum in Sec.~\ref{s: strong}. We again suppose that $\Gamma$
is a connected infinite $C^{4}$ smooth curve straight outside a
compact. For large $\alpha$ the wave function is localized in a
small neighborhood of $\Gamma$ and one can expect that the
reflection and transmission amplitudes which determine the
scattering in the negative part of the spectrum will be determined
in the leading order by the local geometry of the curve.

Let $\mathcal{S}_{\Gamma,\alpha}(\lambda)$ be the on-shell
scattering matrix at energy $\lambda$ for the operator
$H_{\alpha,\Gamma}$, compared to the free dynamics in the leads,
and denote by $\mathcal{S}_{S_\Gamma}(\lambda)$ the corresponding
quantity for the one-dimensional comparison operator $S_\Gamma$ of
Theorem~\ref{infin_asympt}. One conjectures that for a fixed $k\ne
0$ and $\alpha\to\infty$ we have the relation
 $$
\mathcal{S}_{\Gamma,\alpha}\Big(k^2- \frac{1}{4} \alpha^2\Big)\to
\mathcal{S}_{S_\Gamma}(k^{2})\,;
 $$
to prove it and to find the convergence rate one has to show that
the corresponding generalized eigenfunction of $H_{\alpha,\Gamma}$
and that of $S_\Gamma$ multiplied by the transverse eigenfunction
analogous to that in (\ref{gensigma}) converge to each other, and
how fast.

\subsection{More on strong coupling asymptotics} \label{ss: angle}

This brings us to strong-coupling asymptotic results of
Sec.~\ref{s: strong}. One question left open there concerns
manifolds with boundaries; for simplicity let us consider a finite
curve $\Gamma$ which is not closed. As we have pointed out in the
proof of Theorem~\ref{loop2_asympt} the direct use of the
bracketing technique gives the first assertion of the theorem but
they are not precise enough to yield an asymptotic formula. We
conjecture that it is again of the form
$$ \lambda_j(\alpha) = -\frac{1}{4}\,\alpha^2 + \mu_j +
\mathcal{O}(\alpha^{-1} \ln\alpha)\,, $$
where $\mu_j$ is now the $j$-th eigenvalue of the operator given
by the same expression, but with \emph{Dirichlet} boundary
conditions. A way to prove this result would be to consider a
bracketing on prolonged tubular neighborhoods extending to the
distance of order $a$ over the endpoints of $\Gamma$; a trouble to
overcome is that one cannot then separate variables in the leading
order.

All the considerations of Sec.~\ref{s: strong} required a
sufficient smoothness of $\Gamma$. If this is not true the results
are no longer valid. As an example consider again the operator
$H_2(\beta)$ of Sec.~\ref{ss: star}; using a scaling
transformation we find that $\mu_j$ in the above formula has to be
replaced by $-\left(\lambda_j+\frac14 \right)\alpha^2$ where
$\lambda_j$ is the $j$-th eigenvalue of $H_2(\beta)$ corresponding
to $\alpha=1$. We conjecture that if an otherwise smooth $\Gamma$
has one angle equal to $\pi-\beta$, the $j$-th eigenvalue
asymptotics of the corresponding $H_{\alpha,\Gamma}$ is again
$\lambda_j \alpha^2 + o(\alpha^2)$; if there are more angles the
situation is similar but the numbering of the eigenvalues changes
in the appropriate way.

\subsection{Strong-coupling graph limit} So far we have considered
strong coupling for a single curve. If $\Gamma$ is a nontrivial
graph with branchings, one expects a behavior similar to that of a
curve with angles: each vertex will contribute to the asymptotics
below $-\frac12\alpha^2$ by $\lambda_j \alpha^2 + o(\alpha^2)$
where $\lambda_j$'s are now the eigenvalues of the corresponding
leaky star-graph Hamiltonian $H_n(\beta)$ with $n$ being the
vertex degree.

This leads us to the question, whether in the limit
$\alpha\to\infty$ one can get a meaningful expression of the
standard quantum graph type, i.e. an operator on $L^2(\Gamma)$.
The question makes sense, of course, only if we perform a suitable
energy renormalization. The most natural way to do that is to stay
in the vicinity of the ``transverse threshold'', i.e. to subtract
the diverging factor $-\frac14\alpha^2$. If the above conjecture
is valid the limit will be generically trivial, i.e. a graph with
Dirichlet-decoupled edges. There could be nontrivial limits,
however, in cases when the involved family of leaky-graph
Hamiltonians has a threshold resonance (or a singularity which
stays in the vicinity of such a resonance in the limiting
process). The problem is analogous to the squeezing of Dirichlet
fat graphs mentioned in the introduction, cf.~\cite{Po2, CaE, Gr},
and it is likely to be no less difficult. One can also conjecture
that if the reference point is chosen instead as $\lambda
\alpha^2$ with \mbox{$\lambda> -\frac14\alpha^2$} one gets
generically a nontrivial limit with the vertex conditions
determined by the scattering on the appropriate leaky star graph
in analogy with \cite{MV}.

\subsection{More on resonances}

The presence or absence of resonances is one of the most important
features in scattering. In this survey we touched that subject but
not very deeply: we were able to establish the existence of
resonances in the caricature model of Sec.~\ref{ss: line_point}
and we quoted numerical results which suggest that resonances are
present in some other situations. The first problem here is
whether one is able to establish the existence of resonances for a
wider class of operators $H_{\alpha,\Gamma}$ with an infinite
$\Gamma$ having some number of (asymptotically) straight
``leads''.

The question has to be put more precisely. The resonances in
Theorem~\ref{line_point_res} are understood in the common sense as
poles of the analytically continued resolvent. For the model in
question one can check easily \cite{EK6} that they are at the same
time singularities of the scattering matrix. A similar equivalence
between the resolvent and scattering resonances is expected to be
valid for other leaky-graph Hamiltonians as well, and its
verification should accompany the existence proof.

One more problem concerns the method we mentioned in Sec.~\ref{ss:
num}, called usually the \emph{$L^2$--approach} by the physicists,
which allows to identify resonances by spectral methods. As we
have noted the method is rigorously justified for the
one-dimensional potential scattering only \cite{HM} and it is
naturally desirable to find an appropriate formulation and proof
of it in the present situation.

\subsection{More on magnetic leaky graphs}

This is another subject not much explored so far, and also one
where the analogy between leaky graphs and quantum waveguides
sometimes fails, e.g., a waveguide in a homogeneous field has
dominantly a continuous spectrum due to edge states ``skipping''
along the boundary while the spectrum of its leaky-graph
counterpart is a point one being dominated by the Landau levels. A
numerical example worked out in \cite{ETa} using mode matching
indicates that the asymptotic behavior described in
Theorem~\ref{mgloop} can be destroyed if we keep $\alpha$ fixed,
albeit large, and make the magnetic field strong.

In connection to that one can ask, e.g., what will happen with the
curvature-induced bound states of Theorem~\ref{dsexist} if the
system is exposed to a magnetic field \mbox{\`{a} la}
(\ref{mg_Ham}), in particular, whether they will survive an
arbitrarily strong field. In a similar vein, it is possible to ask
what will happen with the curious magnetic transport described in
Theorem~\ref{noedge} if the point interaction array in question is
not straight but ``bent'' and only asymptotically straight.

A much weaker perturbation is represented by local magnetic
fields. Here again one can derive an inspiration from the
waveguide theory: it is known that spectrum of a waveguide with
such a field is, contrary to the non-magnetic case, stable with
respect to small perturbation as a consequence of a Hardy-type
inequality \cite{EKo}. One can ask whether a similar result is
true for leaky curves slightly different from a straight line if a
local magnetic field, regular or singular, is present.

\subsection{Perturbations of periodic graphs}

In addition to the absolute continuity problem mentioned above
periodic graphs pose other questions. The spectrum is expected to
have a gap structure, and in some cases one is able to establish
existence of open gaps as, e.g., in Corollary~\ref{gap}. In such a
case one can ask about the effect of local perturbation to such a
periodic $\Gamma$, in particular, under which conditions they give
rise to eigenvalues in gaps, what is the number of the latter and
their dependence on the perturbation parameters, etc.

\subsection{Absence of embedded eigenvalues}

Another question which arises in connection with the previous problem, but it it is not restricted to perturbations of periodic graphs, concerns embedded eigenvalues. It is well know that in the ``usual'' quantum graphs the unique continuation principle in not valid \cite{KV}, and as a consequence, we encounter frequently situations with eigenvalues embedded into the continuum and associated with compactly supported eigenfunctions. Leaky graphs are different and one naturally asks whether one can establish generally absence of such eigenvalues, and also about possible existence examples if there are any.

\subsection{Random leaky graphs}

Quantum graphs with various sorts of randomness have been recently
an object of intense interest -- see, e.g., \cite{ASW, EHS, KP,
GV} and references therein. In contrast, not much was done about
the analogous problem for leaky graphs which are certainly of
interest because they mix elements of a one-dimensional and
multidimensional behavior. One naturally expects a localization if
either the edge shapes or the coupling constants on them become
random, and there is a numerical evidence supporting these
expectations \cite{ETa}. The truly important question, however,
concerns the existence and properties of a mobility edge in such
systems.

The effects of randomness can be also studied through a point
interaction counterparts to the operators $H_{\alpha,\Gamma}$. If
we take a point-interaction polygon with a magnetic field
mentioned in Sec.~\ref{ss: edge} and randomize the coupling
constant, one can observe numerically how the transport is
destroyed. One can ask therefore what can be proven about the
spectrum of the model from Theorem~\ref{noedge} if the coupling
constants are made random, whether there is a localization and
whether a part of the absolutely continuous spectrum will survive
-- recall that for a similar model with an array replaced by a
rectangular point-interaction lattice a localization was proved in
the low-lying spectral bands \cite{DMP}.

\subsection{Time evolution} \label{ss: time}

So far we spoke about the stationary aspect of the problem, even
when we discussed problems such as scattering. We do not know much
about the way the wave functions evolve in leaky-graph systems.
One can ask, for instance, about smoothing properties analogous to
those of the usual Schr\"odinger operators. In some cases this is
true --- take the example of $H_{\alpha,\Gamma}$ corresponding to
a straight line in the plane --- but a caution is needed since it
is known that the time evolution in systems with singular
interactions can have sometimes rather counterintuitive properties
\cite{EFr2}.

A more specific question is associated with resonances in quantum
graphs. The usual duality between resonances and unstable states
motivates us to ask what happens if we fix a resonant state,
typically an eigenstate of an unperturbed operator associated with
$H_{\alpha,\Gamma}$ embedded into the continuum, at an initial
instant, and ask about the way in which it decays. For the
caricature model of Sec.~\ref{ss: line_point} it can be done
\cite{EIK}; one would welcome to learn more about the decay
processes in physically more interesting cases.

\bigskip

This problem list is in no case exhaustive. If the reader made it
to this point --- my greetings to such a persistent colleague ---
I am sure he or she managed to formulate many additional questions
on the way, and is ready to address them. It remains for me only
to express a good luck wish in such an effort.


\bibliographystyle{amsalpha}

\begin{thebibliography}{A}

\bibitem[ACF03]{ACF}
 A.~Abrams, J.~Cantarella, J.G.~Fu, M.~Ghomi, R.~Howard:
 \textit{Circles minimize most knot energies}, Topology \textbf{42}
 (2003), 381--394.

\bibitem [ASW06]{ASW}
M. Aizenman, R. Sims and S. Warzel: \emph{Absolutely continuous
spectra of quantum tree graphs with weak disorder}, Commun. Math.
Phys. \textbf{264} (2006), 371--389.

\bibitem[AGHH04]{AGHH}
S.~Albeverio, F.~Gesztesy, R.~H\o egh-Krohn, H.~Holden: {\em
Solvable Models in Quantum Mechanics}, 2nd printing, AMS,
Providence, R.I., 2004.

\bibitem[AN00]{AN}
S.~Albeverio, L.~Nizhnik: \emph{Approximation of general
zero-range potentials}, Ukrainian Math. J. {\bf 52} (2000),
664--672.

\bibitem[AGS87]{AGS}
J.-P.~Antoine, F.~Gesztesy, J.~Shabani: \textit{Exactly solvable
models of sphere interactions in quantum mechanics}, J. Phys. A:
Math. Gen. {\bf 20} (1987), 3687--3712.

\bibitem[ABGM91]{ABGM}
Y.~Avishai, D.~Bessis, B.M.~Giraud, G.~Mantica: \emph{Quantum
bound states in open geometries}, Phys. Rev. {\bf B44} (1991),
8028--8034.

\bibitem[BDE03]{BDE}
F.~Bentosela, P.~Duclos, P.~Exner: {\em Absolute continuity in
periodic thin tubes and strongly coupled leaky wires}, Lett. Math.
Phys. {\bf 65} (2003), 75--82.

\bibitem[BS\v{S}00]{BSS} M.S.~Birman, T.A.~Suslina, R.G.~Shterenberg:
\textit{Absolute continuity of the two-dimen\-sional Schr\"odinger
operator with delta potential concentrated on a periodic system of
curves}, Algebra i Analiz \textbf{12} (2000), 140--177; transl. in
St. Petersburg Math. J. \textbf{12} (2001), 535--567.

\bibitem[BCFK06]{BCFK}
G.~Berkolaiko, R.~Carlson, S.~Fulling, P.~Kuchment, eds.:
\textit{Quantum Graphs and Their Applications}, Contemporary
Math., vol.~415, AMS, Providence, R.I., 2006.

\bibitem[BEK\v{S}94]{BEKS} J.F.~Brasche, P.~Exner, Yu.A.~Kuperin,
P.~\v{S}eba: \textit{Schr\"odinger operators with singular
interactions}, J. Math. Anal. Appl. \textbf{184} (1994), 112--139.

\bibitem[BFT98]{BFT}
J.P.~Brasche, R.~Figari, A.~Teta: Singular Schr\"odinger operators
as limits of point interaction Hamiltonians, {\em Potential Anal.}
{\bf 8} (1998), 163--178.

\bibitem[BO07]{BO}
J.F.~Brasche, K.~O\v{z}anov\'{a}: \emph{Convergence of
Schr\"odinger operators}, SIAM J. Math. Anal. {\bf 39} (2007),
281--297.

\bibitem[BT92]{BT}
J.F.~Brasche, A.~Teta: \textit{Spectral analysis and scattering
for Schr\"odinger operators with an interaction supported by a
regular curve}, in ``Ideas and Methods in Quantum and Statistical
Physics'', ed. by S.~Albeverio, J.E.~Fenstadt, H.~Holden,
T.~Lindstr\o m, Cambridge Univ. Press 1992, pp.~197--211.

\bibitem [BEW08]{BEW}
B.M.~Brown, M.S.P.~Eastham, I.G.~Wood: \textit{An example on the
discrete spectrum of a star graph}, in this volume, pp.~??

\bibitem [CE07]{CaE}
C.~Cacciapuoti, P.~Exner: \textit{Nontrivial edge coupling from a
Dirichlet network squeezing: the case of a bent waveguide}, J.
Phys. A: Math. Theor. \textbf{A40} (2007), F511--F523.

\bibitem [CEK04]{CEK}
G.~Carron, P.~Exner, D.~Krej\v{c}i\v{r}{\'\i}k: {\em Topologically
non-trivial quantum layers}, J. Math. Phys. {\bf 45} (2004),
774--784.

\bibitem [Ch99]{Ch}
I.~Chavel, \textit{The Laplacian on Riemannian manifolds},
Proc. ICMS Instructional Conference ``Spectral Theory and
geometry'', Cambridge University Press 1999, pp. 30--75.

\bibitem[ChE03]{ChE2}
T.~Cheon, P.~Exner: {\em Persistent currents due to point
obstacles}, Phys. Lett. {\bf A307} (2003), 209--214.

\bibitem [CE04]{ChE}
T.~Cheon, P.~Exner: \textit{An approximation to $\delta'$
couplings on graphs}, J. Phys. A: Math. Gen. \textbf{37} (2004),
L329--335.

\bibitem [CS98]{CS}
T.~Cheon, T~.Shigehara: \emph{Realizing discontinuous wave
function with renormalized short-range potentials}, Phys. Lett.
{\bf A243} (1998), 111--116.

\bibitem [DMP99]{DMP}
T.C.~Dorlas, N.~Macris, J.V.~Pul\'{e}: \emph{Characterization of
the spectrum of the Landau Hamiltonian with delta impurities},
Commun. Math. Phys. {\bf 204} (1999), 367--396.

\bibitem [Du08]{Du}
P.~Duclos: \textit{On the skeleton method and an aplication to a
quantum scissor}, in this volume, pp.~??

\bibitem [DE95]{DE}
P.~Duclos, P.~Exner: {\em Curvature--induced bound states in
quantum waveguides in two and three dimensions}, Rev. Math. Phys.
{\bf 7} (1995), 73--102.

\bibitem [DEK01]{DEK}
P.~Duclos, P.~Exner, D.~Krej\v{c}i\v{r}{\'\i}k: {\em Bound states
in curved quantum layers}, Commun. Math. Phys. {\bf 223} (2001),
13--28.

\bibitem [EKo05]{EKo}
T.~Ekholm, H.~Kova\v{r}\'{\i}k: \emph{Stability of the magnetic
Schrödinger operator in a waveguide}, Comm. PDE {\bf 30} (2005),
539--565.

\bibitem [Ex96]{Ex1}
P.~Exner: \textit{Weakly coupled states on branching graphs},
Lett.Math.Phys. \textbf{38}(1996),313--320.

\bibitem [Ex01]{Ex4}
P.~Exner: {\em Bound states of infinite curved polymer chains},
Lett. Math. Phys. {\bf 57} (2001), 87--96.

\bibitem [Ex03]{Ex3}
P.~Exner: {\em  Spectral properties of Schr\"odinger operators
with a strongly attractive $\delta$ interaction supported by a
surface}, Proc. of the NSF Summer Research Conference (Mt. Holyoke
2002); AMS ``Contemporary Mathematics" Series, vol.~339,
Providence, R.I., 2003; pp.~25--36.

\bibitem [Ex05a]{Ex5}
P.~Exner: {\em An isoperimetric problem for point interactions},
J. Phys. A: Math. Gen. {\bf 38} (2005), 4795--4802.

\bibitem [Ex05b]{Ex2}
P.~Exner: {\em An isoperimetric problem for leaky loops and
related mean-chord inequalities}, J. Math. Phys. {\bf 46} (2005),
062105

\bibitem [Ex06]{Ex6}
P. Exner: {\em Necklaces with interacting beads: isoperimetric
problems}, Proceedings of the ``International Conference on
Differential Equations and Mathematical Physics'' (Birmingham
2006), AMS ``Contemporary Math" Series, vol.~412, Providence,
R.I., 2006; pp.~141--149.

\bibitem[EF07a]{EFr2}
P.~Exner, M.~Fraas: {\em The decay law can have an irregular
character}, J. Phys. A: Math. Theor. {\bf 40} (2007), 1333--1340.

\bibitem[EF07b]{EFr}
P.~Exner, M.~Fraas: {\em On the dense point and absolutely
continuous spectrum for Hamiltonians with concentric $\delta$
shells}, Lett. Math. Phys. (2007), to appear;
\texttt{arXiv:~0705.1407}

\bibitem[EF07c]{EF}
P.~Exner, R.~Frank: {\em Absolute continuity of the spectrum for
periodically modulated leaky wires in $\mathbb{R}^3$}, Ann.
H.~Poincar\'{e} {\bf 8} (2007), 241--263.

\bibitem[EHL06]{EHL}
P.~Exner, E.M.~Harrell, M.~Loss: {\em Inequalities for means of
chords, with application to isoperimetric problems},
Lett.Math.Phys. {\bf 75} (2006), 225--233; addendum {\bf 77}
(2006), 219

\bibitem[EHS07]{EHS}
P.~Exner, M.~Helm, P.~Stollmann: {\em Localization on a quantum
graph with a random potential on the edges}, Rev. Math. Phys. {\bf
19} (2007), 923--939.

\bibitem[EI01]{EI}
P.~Exner, T.~Ichinose: \textit{Geometrically induced spectrum in
curved leaky wires}, J. Phys. \textbf{A34} (2001), 1439--1450.

\bibitem[EIK07]{EIK}
P. Exner, T. Ichinose, S. Kondej: {\em On relations between stable
and Zeno dynamics in a leaky graph decay model}, Operator Theory:
Advances and Applications, vol.~174, Birk\-h\"auser, Basel 2007;
pp.~21--34.

\bibitem[EJK99]{EJK}
P.~Exner, A.~Joye, H.~Kova\v{r}\'{\i}k: {\em Edge currents in the
absence of edges}, Phys. Lett. {\bf A264} (1999), 124--130.

\bibitem[EK02]{EK2}
P.~Exner, S.~Kondej: \textit{Curvature-induced bound states for a
$\delta$ interaction supported by a curve in $\mathbb{R}^3$}, Ann.
H.~Poincar\'{e} \textbf{3} (2002), 967--981.

\bibitem[EK03]{EK}
P.~Exner, S.~Kondej: \textit{Bound states due to a
strong $\delta$ interaction supported by a curved surface}, J.
Phys. A: Math Gen. \textbf{36} (2003), 443--457.

\bibitem[EK04a]{EK3}
P.~Exner, S.~Kondej: {\em Strong-coupling asymptotic expansion for
Schr\"odinger operators with a singular interaction supported by a
curve in $\mathbb{R}^3$}, Rev. Math. Phys. {\bf 16} (2004),
559--582.

\bibitem[EK04b]{EK6}
P.~Exner, S.~Kondej: {\em Schr\"odinger operators with singular
interactions: a model of tunneling resonances}, J. Phys. {\bf A37}
(2004), 8255--8277.

\bibitem[EK05]{EK5}
P.~Exner, S.~Kondej: {\em Scattering by local deformations of a
straight leaky wire}, J. Phys. {\bf A38} (2005), 4865--4874.

\bibitem[EK07]{EK4}
P.~Exner, S.~Kondej: \emph{Hiatus perturbation for a singular
Schr\"odinger operator with an interaction supported by a curve in
$\mathbb{R}^3$}, a paper in preparation

\bibitem[EKr01]{EKr}
P.~Exner, D.~Krej\v{c}i\v{r}{\'\i}k: {\em Bound states in mildly
curved layers}, J. Phys. {\bf A34} (2001), 5969--5985.

\bibitem[ENZ01]{ENZ}
P.~Exner, H.~Neidhardt, V.A.~Zagrebnov: {\em Potential
approximations to $\delta'$: an inverse Klauder phenomenon with
norm-resolvent convergence}, Commun. Math. Phys. {\bf 224} (2001),
593--612.

\bibitem[EN01]{EN2}
P.~Exner, K.~N\v{e}mcov\'{a}: {\em Bound states in point
interaction star graphs}, J. Phys. A: Math. Gen. {\bf 34} (2001), 7783--7794.

\bibitem[EN03]{EN1}
P.~Exner, K.~N\v{e}mcov\'{a}: {\em Leaky quantum graphs:
approximations by point inter\-action Hamiltonians}, J. Phys. A:
Math Gen. {\bf 36} (2003), 10173--10193.

\bibitem[EP05]{EP1}
P.~Exner, O.~Post: \textit{Convergence of spectra of graph-like
thin manifolds}, J. Geom. Phys. {\bf 54} (2005), 77--115.

\bibitem[EP07]{EP2}
P.~Exner, O.~Post: \textit{Convergence of resonances on thin branched
quantum wave guides}, J. Math. Phys. {\bf 48} (2007), 092104

\bibitem[E\v{S}89]{ES}
P.~Exner, P.~\v{S}eba: \textit{Bound states in curved quantum
waveguides}, J. Math. Phys. {\bf 30} (1989), 2574--2580.

\bibitem[ET04]{ETa}
P.~Exner, M.~Tater: {\em Spectra of soft ring graphs}, Waves in
Ran\-dom Media {\bf 14} (2004), S47--60.

\bibitem[ETu07]{ET}
P.~Exner, O.~Turek: \textit{Approximations of singular vertex
couplings in quantum graphs}, Rev. Math. Phys. {\bf 19} (2007), 571--606.

\bibitem [EY01]{EY2}
P.~Exner, K.~Yoshitomi, \textit{Band gap of the Schr\"odinger
operator with a strong $\delta$-interaction on a periodic curve},
Ann. H.~Poincar\'e \textbf{2} (2001), 1139--1158.

\bibitem [EY02a]{EY1} P.~Exner, K.~Yoshitomi, \textit{Asymptotics of
eigenvalues of the Schr\"odinger operator with a strong
$\delta$-interaction on a loop}, J. Geom. Phys. \textbf{41}
(2002), 344--358.

\bibitem [EY02b]{EY3}
P.~Exner, K.~Yoshitomi, \textit{Persistent currents
for 2D Schr\"odinger operator with a strong $\delta$-interaction
on a loop}, J. Phys. A: Math. Gen. \textbf{35} (2002), 3479--3487.

\bibitem [EY03]{EY4}
P.~Exner, K.~Yoshitomi: {\em Eigenvalue asymptotics for the
Schr\"odinger operator with a $\delta$-inter\-action on a
punctured surface}, Lett. Math. Phys. {\bf 65} (2003), 19--26;
erratum {\bf 67} (2004),~81--82.

\bibitem[FKu96]{FK}
A.~Figotin, P.~Kuchment: \textit{Band-gap structure of spectra
of periodic dielectric and acoustic media I, II}, SIAM J. Appl.
Math. \textbf{56} (1996), 68--88, 1561--1620.

\bibitem[FKu98]{FK2}
A.~Figotin, P.~Kuchment: \textit{Spectral properties of classical
waves in high-contrast periodic media}, SIAM J. Appl. Math.
\textbf{58} (1998), 683--702.

\bibitem[FKl04]{FKl}
N.~Filonov, F.~Klopp: \textit{Absolute continuity of the spectrum
of a Schr\"odinger operator with a potential which is periodic in
some directions and decays in others}, Doc. Math. \textbf{9}
(2004), 107--121; erratum ibidem 135--136.

\bibitem[Fr03]{Fr1}
R.L.~Frank: \textit{On the scattering theory of the Laplacian with
a periodic boundary condition. I. Existence of wave operators},
Doc. Math. \textbf{8} (2003), 547--565.

\bibitem[Fr06]{Fr2} R.L.~Frank:
\textit{On the Laplacian in the halfspace with a periodic boundary
condition},  Ark. Mat. \textbf{44} (2006), 277--298.

\bibitem[FS04]{FS}
R.L.~Frank, R.G.~Shterenberg: \textit{On the scattering theory of
the Laplacian with a periodic boundary condition. II. Additional
channels of scattering}, Doc. Math. \textbf{9} (2004), 57--77.

\bibitem[FS07]{FS2}
R.L.~Frank, R.G.~Shterenberg: \textit{On the spectrum of partially
operators}, in ``Operator Theory, Analysis and Mathematical
Physics'', Oper. Theory. Anal. Appl. \textbf{174} (2007), 35--50.

\bibitem[FW93]{FW}
M.~Freidlin, A.~Wentzell: \textit{Diffusion processes on graphs
and the averaging principle}, Ann. Prob. \textbf{21} (1993),
2215--2245.

\bibitem[FKr07]{FKr}
P.~Freitas, D.~Krej\v{c}i\v{r}\'{\i}k: \emph{Location of the nodal
set for thin curved tubes}, Indiana Univ. Math. J., to appear;
\texttt{math.SP/0602470}

\bibitem[Gr07]{Gr}
D.~Grieser: \textit{Spectra of graph neighborhoods and
scattering}, \texttt{arXiv:0710.3405v1 [math.SP]}

\bibitem[GV07]{GV}
M.J.~Gruber, I.~Veseli\'{c}: \emph{The modulus of continuity of
Wegner estimates for random Schrödinger operators on metric
graphs}, \texttt{arXiv:0707.1486 [math.SP]}

\bibitem[HM00]{HM}
G.A.~Hagedorn, B.~Meller: \emph{Resonances in a box}, J. Math.
Phys. {\bf 41} (2000), 103--117.

\bibitem[Ha82]{Ha}
B.I.~Halperin: \emph{Quantized Hall conductance, current-carrying
edge states, and the existence of extended states in a
two-dimensional disordered potential}, Phys. Rev. {\bf B25}
(1982), 2185--2190.

\bibitem[He89]{He}
J.~Herczynski: \emph{On Schr\"odinger operators with a
distributional potential}, J. Operator Theory {\bf 21} (1989),
273--295.

\bibitem[Il92]{Il}
A.~Il'in: {\em Matching of Asymptotic Expansions of Solutions of
Boundary Value Problems}, Translations of Mathematical Monographs,
vol. 102, AMS, Providence, R.I., 1992.

\bibitem[Ka76]{Ka}
T.~Kato: {\em Perturbation Theory for Linear Operators}, 2nd
edition, Springer, Berlin 1976.

\bibitem [Kli78]{Kli} W.~Kligenberg, \textit{A Course in Differential Geometry},
Springer Verlag, New York 1978.

\bibitem[KP07]{KP}
F.~Klopp, K.~Pankrashkin: \emph{Localization on quantum graphs
with random vertex couplings}, \texttt{arXiv:0710.3065}

\bibitem[Ku93]{Ku3}
P.~Kuchment: \textit{Floquet Theory For Partial Differential
Equations}, Birkh\"auser Verlag, Basel 1993.

\bibitem[Ku01]{Ku}
P.~Kuchment: \textit{The mathematics of photonic crystals},
Chap.~7 in ``Mathematical Modeling in Optical Science'' (Gang Bao,
Lawrence Cowsar, and Wen Masters, eds.), Frontiers in Applied
Mathematics, vol.~22, SIAM 2001; pp.~207--272.

\bibitem[Ku02]{Ku2}
P.~Kuchment: \textit{Graph models of wave propagation in thin
structures}, Waves in Random Media \textbf{12} (2002), R1--24.

\bibitem[KK98]{KK}
P.~Kuchment, L.~Kunyansky: \textit{Spectral properties of
high-contrast band-gap materials and operators on graphs},
Experimental Math. \textbf{8} (1998), 1--28.

\bibitem[KK02]{KK2}
P.~Kuchment, L.~Kunyansky: \textit{Differential operators on
graphs and photonic crystals}, Adv. Comp. Math. \textbf{16}
(2002), 263--290.

\bibitem[KV06]{KV}
P.~Kuchment, B.~Vainberg: \textit{ On the structure of eigenfunctions corresponding to embedded eigenvalues of locally perturbed periodic graph operators}, Commun. Math. Phys. \textbf{268} (2006), 673--686.

\bibitem[KZ01]{KZ}
P.~Kuchment, H.~Zeng: \textit{Convergence of spectra of mesoscopic
systems collapsing onto a graph}, J. Math. Anal. Appl.
\textbf{258} (2001), 671--700.

\bibitem[KZ03]{KZ2}
P.~Kuchment, H.~Zeng: \textit{Asymptotics of spectra of Neumann
Laplacians in thin domains}, in ``Advances in Differential
Equations and Mathematical Physics,'' (Yu. Karpeshina, G. Stolz,
R. Weikard, Y. Zeng, eds.), Contemporary Mathematics, vol.~387,
AMS 2003; pp.~199--213.

\bibitem[Lu66]{Lu}
G.~L\"uk\H{o}: \textit{On the mean lengths of the chords of a
closed curve}, Israel J. Math. \textbf{4} (1966), 23--32.

\bibitem[MDS84]{MDS}
A.H.~MacDonald, P.~St\v{r}eda: \emph{Quantized Hall effect and
edge currents}, Phys. Rev. {\bf B29} (1984), 1616--1619.

\bibitem[MV07]{MV}
S.~Molchanov, B.~Vainberg: \textit{Scattering solutions in a
network of thin fibers: small diameter asymptotics}, Commun. Math.
Phys. \textbf{273} (2007), 533--559.

\bibitem[O\v{z}06]{Oz}
K.~O\v{z}anov\'{a}: \emph{Approximation by point potentials in a
magnetic field}, J. Phys. A: Math. Gen. {\bf 39} (2006),
3071--3083.

\bibitem[PP04]{PP}
O.~Petri, L.~P\"aiv\"arinta: {\em Mellin operators and
pseudodifferential operators on graphs}, Waves in Ran\-dom Media
{\bf 14} (2004), S129--142.

 \bibitem[Pos01]{Pos1}
A.~Posilicano: \textit{A Krein-Like Formula for Singular
Perturbations of Self-Adjoint Operators and Applications}, J.
Funct. Anal. {\bf 183} (2001), 109--147.

\bibitem[Pos04]{Pos2}
A.~Posilicano: \textit{Boundary triples and Weyl functions for
singular perturbations of self-adjoint operator}, Meth. Funct.
Anal. Topol. {\bf 10} (2004), 57--63.

\bibitem[Po05]{Po2}
O.~Post: \textit{Branched quantum wave guides with Dirichlet
boundary conditions: the decoupling case}, J. Phys. A: Math. Gen.
\textbf{38} (2005), 4917--4931.

\bibitem[Po06]{Po1}
O.~Post: \textit{Spectral convergence of non-compact
quasi-one-dimensional spaces}, Ann. H. Poincar\'{e} \textbf{7}
(2006), 933--973.

\bibitem[RSi79]{RSi}
M.~Reed and B.~Simon: {\em Methods of Modern Mathematical Physics,
III.~Scat\-tering Theory}, Academic Press, New York 1979.

\bibitem[RS01]{RS}
J.~Rubinstein, M.~Schatzmann: \textit{Variational problems on
multiply connected thin strips, I.~Basic estimates and convergence
of the Laplacian spectrum}, Arch. Rat. Mech. Anal. \textbf{160}
(2001), 271--308.

\bibitem[Sa01]{Sa}
T.~Saito: \textit{Convergence of the Neumann Laplacian on
shrinking domains}, Analysis \textbf{21} (2001), 171--204.

\bibitem[Sha88]{Sha}
J.~Shabani: \textit{Finitely many delta interactions with supports on
concentric spheres}, J. Math. Phys. {\bf 29} (1988), 660--664.

\bibitem[S\v{S}01]{SS}
T.A.~Suslina, R.G.~Shterenberg: \textit{Absolute continuity of the
spectrum of the Schr\"odinger operator with the potential
concentrated on a periodic system of hypersurfaces}, Algebra i
Analiz \textbf{13} (2001), 197--240.
















\end{thebibliography}

\end{document}